\documentclass{article}[12pt]
\usepackage[utf8]{inputenc}
\usepackage[T1]{fontenc}
\usepackage{xcolor}
\usepackage{graphicx}
\usepackage{hyperref}
\usepackage{subcaption}
\usepackage{tikz}

\hypersetup{%
colorlinks=true,
breaklinks=true,
linkcolor=red,
anchorcolor=black,
citecolor=brown,
filecolor=blue,
menucolor=red,
urlcolor=red,
}
\usepackage[round,longnamesfirst]{natbib}
\usepackage{mathtools, amsfonts, amssymb,amsthm,stmaryrd,amsmath}
\mathtoolsset{showonlyrefs}
\usepackage{enumerate}
\usepackage{comment}
\usepackage{ifthen}
\usepackage{mathrsfs}
\usepackage{enumerate}
 \usepackage{filecontents}

\newtheorem{assumpA}{Assumption}
\newtheorem{definition}{Definition}

\newcommand{\EE}{\mathbb{E}}
\newcommand{\EEMT}{\mathbb{E}_{M,T}}
\newcommand{\PP}{\mathbb{P}}

\newcommand{\N}{\||}

\newtheorem{theorem}{Theorem}
\newtheorem{corollary}{Corollary}

{%
\medskip
\noindent\textbf{Assumptions.}
  \begin{enumerate}[({H}1)]%
  \setcounter{enumi}{\value{assumptionH}}%
}{%
  \setcounter{assumptionH}{\value{enumi}}%
  \end{enumerate}
}

\newcommand{\assrefH}[1]{(\hyperref[#1]{H\ref{#1}})}

\newcommand{\Mmu}{\mathfrak{m}}

\newtheorem{thApp}{Theorem A.\!\!}
\newtheorem{lemApp}{Lemma A.\!\!}

\mathtoolsset{showonlyrefs}
\usepackage{xr}
\title{Adaptive functional principal components analysis}
\author{Sunny G.W. Wang\footnote{Ensai, CREST - UMR 9194, France; sunny.wang@ensai.fr} \qquad
Valentin Patilea\footnote{Ensai, CREST - UMR 9194, France; valentin.patilea@ensai.fr}
\qquad
Nicolas Klutchnikoff\footnote{Univ Rennes, IRMAR - UMR 6625, France; Nicolas.klutchnikoff@univ-rennes2.fr}}
\date{October 22, 2024}

\begin{document}

\maketitle

\begin{abstract}
Functional data analysis almost always involves smoothing discrete observations into curves, because they are never observed in continuous time and rarely without error. Although smoothing parameters affect the subsequent inference, data-driven methods for selecting these parameters are not well-developed, frustrated by the difficulty of using all the information shared by curves while being computationally efficient. On the one hand, smoothing individual curves in an isolated, albeit sophisticated way, ignores useful signals present in other curves. On the other hand, bandwidth selection by automatic procedures such as cross-validation after pooling all the curves together quickly become computationally unfeasible due to the large number of data points. In this paper we propose a new data-driven, adaptive kernel smoothing, specifically tailored for functional principal components analysis through the derivation of sharp, explicit risk bounds for the eigen-elements. The minimization of these quadratic risk bounds provide refined, yet computationally efficient bandwidth rules for each eigen-element separately. Both common and independent design cases are allowed.  Rates of convergence for the estimators are derived. 
An extensive simulation study, designed in a versatile manner to closely mimic the characteristics of real data sets supports our methodological contribution. An illustration on a real data application is provided.



\end{abstract}


\section{Introduction}\label{introduction}

\subsection{Motivation}
The advent of modern data collection mechanisms, exemplified by the inclusion of sensors, has given rise to a fascinating array of intricate functional data sets. These data sets often exhibit rough trajectories, characterized by varying degrees of smoothness that challenge the conventional framework of functional data analysis (FDA). Traditional FDA methods have long focused on smooth sample paths, tacitly assuming knowledge of their regularity.
However, the current big data landscape calls for adaptive methodologies that can accommodate the idiosyncrasies inherent in these complex functional data sets. While the estimation of mean, covariance, and regression functions has witnessed some advancements in adaptive techniques, the exploration of adaptive functional principal component analysis (FPCA) remains relatively unexplored.
FPCA, with its capacity to extract salient modes of variation from curves, continues to hold a central position in the pantheon of dimension reduction techniques. Frequently employed as a preliminary step in downstream modeling endeavors such as regression and classification, FPCA has historically been executed through eigen-analysis of estimated covariance functions. In densely sampled scenarios, the `smooth-first-then-estimate' approach dominates, while in sparsely sampled situations, `weighting schemes' that pool all points together before estimation tend to prevail. These prevalent strategies assume knowns degree and patterns of smoothness, rendering them ill-suited for complex functional data sets characterized by heterogeneous degrees of smoothness and motivating the need for more flexible frameworks. Furthermore, in the context of FPCA, the notion of adaptivity goes beyond considering only the regularity of the underlying curves, as it also encompasses the adaptivity to each principal component and eigenvalue of the covariance. 

Let us  review a few of the most important contributions
in FPCA. \cite{Bosq2000LinearApplications} provided theoretical grounds  in the ideal fully observed, noiseless case. See also  \cite{Horvath2012}. For the practical aspects, 
the book of \cite{Ramsay2005} is a landmark reference. The active research of the recent years provided advances in more realistic data settings where the curves are contaminated by noise and necessarily subjected to discretization errors. When the data curves admit derivatives up to some known order, \cite{HallMuelle2006}  established $L^2$ optimal rates of convergence for the principal components, using a covariance function estimated by pooling points together and applying a local linear smoother, while \cite{Li2010}  obtained the uniform rates.   \cite{Zhang2016} further refined the convergence results for the  pooling observations approach. \cite{benko2009} propose alternative principal components estimation using a
well-known duality relation between row and column spaces of a data matrix, and kernel smoothing. However, the existing body of methods is largely silent on the practical issue of bandwidth selection, and usually requires some mix of ad-hoc judgment and computationally intensive tools such as cross-validation. For example, in \cite{Zhang2016}, one recommendation is to use a mix of different cross-validation methods, and select the bandwidth that obtains the best fit or is the most interpretable. 
Let us end our brief review by mentioning \cite{Hall2006c}, who first proposed asymptotic expansions for eigen-elements associated to the empirical covariance estimator and their corresponding rates of convergence. Our new approach relies on this type of expansion.

This work aims to develop an FPCA algorithm that meets four important requirements: (i) the algorithm should be flexible enough to adapt to complex functional data sets with irregular sample paths of possibly varying smoothness,  often occurring for instance in energy, meteorology, and medical applications; (ii) it should effectively utilize the replication structure of functional data by incorporating signals from other curves; (iii) be computationally simple and efficient to compute; and (iv) be tailored for each eigen-element of the covariance operator. The last property stems from risk bounds and error rates that we derive, and are expressed as explicit functions of the bandwidth.
Our contribution is synthesized  in the formulation of such an FPCA algorithm, made attainable through a refined plug-in bandwidth rule that minimizes  explicit, sharp quadratic risk bounds that are easy to implement. Our  risk bounds, more precisely the squared bias terms in the bounds, importantly feature the regularity of the process generating the true sample paths, which are usually unknown in practice, motivating the need for methods that automatically adapt to this regularity that typically governs the rate of convergence. Instead of following the direction of more traditional adaptive methods, such as block thresholding, we obtain adaptation by building upon the work of \cite{Golovkine2022}, and explicitly estimating the local regularity of the process generating the data curves instead. The estimation of this regularity is made possible due to the replication nature of functional data. We show that the rates of convergence are governed through the integrals of the local regularity over its sampling domain, and in particular, is driven by the lowest regularity. Contrary to the conventional wisdom of a universal smoothing method for all FDA tasks, we propagate the findings by \cite{carroll2013} and \cite{Golovkine2021}, by showing that for the purposes of FPCA, the optimal bandwidth for smoothing curves is generally different from that of optimal curve recovery, or mean and covariance function estimation. Moreover, each principal component requires a different optimal bandwidth. The theory reveals that the difference between the optimal bandwidths for different principal components comes from the constants multiplying the terms in the quadratic risk bounds. However, like it was established several decades ago for the kernel smoothing for regression curves, the constants matters in applications. We finally want to emphasize that our adaptive approach based on the regularity of the process generating the data curves, requires the number of observation points on each curve to be sufficiently large. The estimation of the process regularity cannot be performed in the case of extremely sparse designs with very few observations per curve, as considered by \cite{HallMuelle2006}.

Our general methodology and the FPCA algorithm are presented in Section \ref{sec_main_metho}. It applies to discretely observed functional data, at random (independent design) or fixed (common design) points. Moreover, the noise is allowed to be heteroscedastic. The justification of our risk bounds and their rates of convergence, as well as the rates of the data-driven bandwidths obtained by the algorithm, 
are provided in Section \ref{sec:th_grnd}. The conclusions of an extensive empirical study are reported in Section \ref{sec:simu-sec}. They are obtained with the newly built \texttt{R} package \texttt{FDAdapt}. For comparison purposes, we use simulated data, but the data are mimicking the features of a real power consumption dataset. We also build a general purpose simulator, which we consider of independent interest, as it can match arbitrary process mean, covariance functions and noise variance situations. This type of device is an effective tool for comparing different approaches in various realistic setups. The simulations illustrate the good performance of our method, and  reasonable computation times. The Appendix gathers the assumptions, and the theoretical statements  behind the local regularity estimation, a new result of independent interest. The Appendix also contains the sketch of the proofs of the results in Section \ref{sec:th_grnd}. The technical details are relegated to the Supplementary Material, where we also present additional simulation results in several different setups.


\section{Methodology}\label{sec_main_metho}
\subsection{The data and curves reconstruction}
Let $X=\{X_t\}_{t \in \mathcal{T}}$ be a stochastic process defined on a compact interval $\mathcal{T}$, say $[0, 1]$. The mean and covariance functions of the process $X$ are given by
\begin{equation}
    \mu(t) = \EE(X_t) \quad \text{and} \quad \Gamma(s,t) = \gamma(s,t) - \mu(s)\mu(t) \quad \text{with} \quad   \gamma(s,t) = \EE(X_s X_t),\qquad s,t\in\mathcal T.
\end{equation}
Let $\lambda_1\geq \lambda_2\geq \ldots\geq 0$ be the ordered eigenvalues of the covariance operator defined by $\Gamma$, and  $\psi_j$  be orthonormal eigenfunctions associated to $\lambda_j$, $j\geq 1$.

The data are collected from $N$ independent realizations $X^{(1)},\ldots X^{(i)},\ldots X^{(N)}$ of $X$. The observed time points along each curve $i$ are denoted $T_m^{(i)}$, they belong to $\mathcal T$ and are indexed by $1 \leq m \leq M_i$. For the sake of readability, we  use the notation $X^{(i)}(T_m^{(i)})$ for $X^{(i)}_t$ with $t=T_m^{(i)}$. Each curve is contaminated by noise, due to  independent, centered errors that are possibly heteroscedastic. Formally, for each $1\leq i \leq N$, the  observations associated to the sample path $X^{(i)}$ are the pairs $(Y_m^{(i)}, T_m^{(i)}) \in \mathbb{R} \times \mathcal{T}$ where 
\begin{equation}\label{data}
    Y_m^{(i)} = X^{(i)}(T_m^{(i)}) + 
         \sigma(T_m^{(i)})e_m^{(i)},\qquad 1 \leq m \leq M_i. 
\end{equation}
The total number  of observed pairs $(Y_m^{(i)}, T_m^{(i)})$ is $M_1+\ldots+M_N$.
The $e_m^{(i)}$'s are independently and identically distributed (iid) centered variables with unit variance. The errors' conditional variance function $\sigma^2(t)$ is bounded and unknown. In the independent design case we assume that the  $M_i$ are iid and generated from a positive integer variable  $M$, with   expectation $\mathfrak{m}$.   Moreover, we assume that the sampling points $T_m^{(i)}$ are iid and generated from $T\in\mathcal T$, and that  $X,M$ and $T$ are mutually independent. In the common design case,  $M_i= \mathfrak{m}$  and the  observed time points $T_1^{(i)},\ldots, T_{\mathfrak m} ^{(i)}$ are non random and the same for all $i$. Without loss of generality, whatever the design is, we assume that for each $i$, $T_1^{(i)},\ldots, T_{M_i} ^{(i)}$ are increasingly ordered. 

In order to proceed with  FPCA, we have to estimate $\Gamma(s,t)$ for any pair $(s,t)$. In the ideal situation where the curves $X^{(i)}$ are observed at any $t\in\mathcal T$ without error, the natural estimates of  $\lambda_j$ and   $\psi_j$ are the empirical ones, that are   obtained from the empirical covariance function
\begin{equation}\label{eq_def_empG}
	\widetilde \Gamma (s,t) = \frac{1}{N} \sum_{i=1}^{N} X^{(i)}_s X^{(i)}_t - \widetilde \mu(s) \widetilde \mu(t) \qquad \text{ where } \qquad  \widetilde \mu(t)  = \frac{1}{N} \sum_{i=1}^{N} X^{(i)}_t, \quad s,t\in\mathcal T.
\end{equation} 
Let    $\widetilde \lambda_j$ and  $\widetilde \psi_j$ be the empirical estimators of the eigenvalue $\lambda_j$ and  eigenfunction $\psi_j$, respectively, that means
$$
\int_{\mathcal T} \widetilde \Gamma(s,t) \widetilde  \psi_j(s)ds =  \widetilde \lambda_j\widetilde \psi_j(t), \quad \forall t\in\mathcal T,  
\qquad   \int_{\mathcal T}  \widetilde \psi_j (t)  \widetilde \psi_{j^\prime} (t) dt = \mathbf{1}\{j=j^\prime\},\quad \forall j,j^\prime \geq 1.
$$
Here, $\mathbf{1}\{.\}$ denote the indicator function.
The properties of $\widetilde \lambda_j$ and  $\widetilde \psi_j$ have been extensively studied, see for instance \cite{Horvath2012}. Herein, for each estimated eigenfunction we adopt the standard convention and consider the version having positive inner product with the true eigenfunction. 

In practice, the curves $X^{(i)}$ are not observed and the estimate of $\Gamma(s,t)$ has to be built using  the data points $(Y_m^{(i)}, T_m^{(i)})$. A convenient way to proceed with real data is the so-called `first smooth, then estimate' approach:
first, using some nonparametric  method, build an estimate $\widehat X^{(i)}$ for each curve $X^{(i)}$ separately; next, build a version $\widehat \Gamma$ of empirical covariance  with the reconstructed curves $\widehat X^{(i)}$ replacing the true $X^{(i)}$'s. The formal definition of $\widehat \Gamma$ is provided in Section \ref{sec:mean-cov-estim}.
 We follow this simple idea and use a linear kernel smoother for each curve $X^{(i)}$, that is
\begin{equation}\label{eq:smoothed_curves_x}
	\widehat X^{(i)}_t = \widehat X^{(i)}_t(h) = \sum_{m=1}^{M_i} W_m^{(i)}(t;h) Y_m^{(i)}, \qquad t\in\mathcal T, \;1 \leq i \leq N,
\end{equation}
where the weights $W_m^{(i)}(t;h)$ are that of Nadaraya-Watson's or local polynomial smoothing. The weights depend on the observed time points $T_m^{(i)}$ and a bandwidth $h$. Let $\widehat \lambda_j$ and $\widehat \psi_j $, $j\geq 1$, denote the eigen-elements obtained from $\widehat \Gamma$. They depend on the bandwidth $h$ and the key issue is the choice of $h$.

\subsection{Risk bounds and sample paths regularity}

The goal is to construct adaptive estimates $\widehat \lambda_j$ and $\widehat \psi_j $ of the eigenvalues $\lambda_j$ and eigenfunctions $\psi_j$, respectively. We consider the quadratic risks associated to the estimators $\widehat \lambda_j$ and $\widehat \psi_j $, that are
\begin{equation}\label{risk_def1}
	\mathcal{R}_N(\widehat \lambda_j ;h) := \EE \left[\left\{\widehat \lambda_j - \lambda_j \right\}^2 \right] \quad \text{and} \quad 	\mathcal{R}_N(\widehat \psi_j;h) := \EE \left\|\widehat \psi_j - \psi_j \right\|_2^2 ,
\end{equation}
respectively. Here, $\|\cdot\|_2$ denotes the $L^2-$norm.  The risks depend on the bandwidth and the challenge is to find a way to select the bandwidths which minimize them.  
By the inequality $(a+b)^2 \leq 2a^2+2b^2$, 
 \begin{equation}\label{ry_dec}
 	\mathcal{R}_N(\widehat \lambda_j ;h) \leq 2\EE \left[\left\{\widehat \lambda_j - \widetilde \lambda_j \right\}^2 \right] + 2\EE \left[\left\{\widetilde \lambda_j - \lambda_j \right\}^2 \right],
 \end{equation}
 and 
 \begin{equation}\label{ry_dec2}
 	\mathcal{R}_N(\widehat \psi_j;h)  \leq 2\EE \left\|\widehat \psi_j - \widetilde \psi_j \right\|_2^2 + 2\EE \left\|\widetilde \psi_j - \psi_j \right\|_2^2.
 \end{equation}
It is well-known that the infeasible empirical estimators $\widetilde \lambda_j$ and  $\widetilde \psi_j$ converge at the parametric rate  $O_{\mathbb{P}}(N^{-1/2})$; see \cite{Horvath2012}. On the other hand, no estimator of the eigen-elements is expected to converge faster than the ones obtained in the ideal situation where the curves $X^{(i)}$ are  observed at any time point without error. Therefore, minimizing the mean squared differences 
 \begin{equation}\label{eq:our_tilde}
\EE \left[\left\{\widehat \lambda_j - \widetilde \lambda_j \right\}^2 \right] \quad \text{ and } \quad \EE \left\|\widehat \psi_j - \widetilde \psi_j \right\|_2^2 ,
 \end{equation}
with respect to the bandwidth, would yield rate optimal estimators of $ \lambda_j$ and $ \psi_j $, respectively. We show that the  mean squared differences in \eqref{eq:our_tilde} can be conveniently bounded when the covariance function is estimated using the `first smooth, then estimate' approach. We are then able to define simple, data-driven bandwidth rules for the kernel smoothing of the curves, to be used next for building covariance estimates, from which the adaptive  estimators $\widehat \lambda_j$ and $\widehat \psi_j $ are derived.

A key element for deriving workable risk bounds, is the local regularity of the process $X$. 
For the sake of simplicity, we here focus on the case where the sample paths of $X$ are non-differentiable. 
There is now extensive evidence that many functional datasets can be reasonably considered as being generated by irregular sample paths $X^{(i)}$. See, for instance, \cite{impact20}, \cite{Mohammadi2022}, \cite{Mohammadi2021} and \cite{Golovkine2022} for examples. Another example of a real data set is studied in Section \ref{sec:simu-sec}.  In Section \color{blue}II.3 \color{black} in the Supplement we discuss the extension of our approach to FDA with smooth sample paths. In the case where $X$ has  non-differentiable (irregular) sample paths, we assume that functions $H:\mathcal T \rightarrow (0,1)$ and  $L:\mathcal T \rightarrow (0,\infty)$ exist such that, for any $t\in \mathcal T$,  
\begin{equation}\label{eq:def_lr1x}
	\EE\left[ (X_u-X_v)^2\right] \approx L^2_t |u-v|^{2H_t},
\end{equation}
when $u\leq t\leq v$ lie in a small neighborhood of $t$. The formal definition behind \eqref{eq:def_lr1x} is provided in Section~\ref{sec:def_loc_reg_x}. A similar notion, called local intrinsic stationarity, was proposed by \cite{hsing2016}. 
The function $H$ provides the local Hölder exponent, while the function $L$ gives the local Hölder constant. They are both allowed to depend on $t$ in order to allow for curves with general patterns, in particular with varying regularity over $\mathcal T$. The function $H$  is also connected to the rate of decay of eigenvalues. For illustration, consider $H\in(0,1)$ is constant. If, for some $1<\nu <3$, the eigenvalues $\lambda_j$  have polynomial decrease rate $\lambda_j \sim j^{-\nu}$, $j\geq 1$, then, under mild conditions, $H$ is constant and equal to $(\nu-1)/2$. For instance, for the Brownian motion $H \equiv 0.5$ and  $\nu = 2$. However, our framework is more general as we do not impose specific rates of decrease for the eigenvalues. 

We show that if $H$ and $L$ are given, sharp bounds for the risk of kernel-based estimates of the eigenvalues and eigenfunctions can be derived. Replacing $H$ and $L$ by their nonparametric estimates in the risk bound of an eigen-element, yields an easy to minimize criterion in order to obtain a data-driven bandwidth adapted to the eigen-element. Building the covariance estimate $\widehat{\Gamma}$ as in  \eqref{cov_est_corr} below with this bandwidth, yields the eigen-element optimal estimate. 

Let us briefly describe the rationale behind the nonparametric estimators of $H$ and $L$. Let  
\begin{equation}\label{def_theta_uv_x}
	\theta(u,v) := \EE\left[ (X_u -  X_v)^{2} \right], \quad u,v\in\mathcal T,
\end{equation}
and consider two points $t_1,t_3\in\mathcal T$ such that $t$ belongs to the interval defined by $t_1$ and $t_3$. Let $t_2 = (t_1+t_3)/2$.
Then, using \eqref{eq:def_lr1x}, it easy to see that 
\begin{equation}\label{eq:tilde-Hd_x}
	\widetilde H_{t} =  \frac{\log(\theta(t_1,t_3)) - \log(\theta(t_1,t_2))}{2\log(2)} \approx H_t \quad \text{ and } \quad  
		\widetilde L_{t} = \frac{\sqrt{\theta(t_1,t_3)} }{|t_1-t_3|^{	\widetilde H_{t} }}\approx L_t, \quad \text{provided $|t_1-t_3|$ is small.}
\end{equation}
 The estimators $\widehat H_t$ and $\widehat L_t$ of $H_t$ and $L_t$ are obtained by simply plugging nonparametric estimators of $\theta(\cdot,\cdot)$ into the expressions of the proxies $\widetilde H_{t}$ and $\widetilde L_{t}$. Details are provided in Section \ref{sec:algorithm}.

\subsection{Covariance estimator}\label{sec:mean-cov-estim}
We now provide the formal definition of our covariance  estimator $\widehat \Gamma$ for a given bandwidth $h$. The estimator is built using kernel estimators as in \eqref{eq:smoothed_curves_x}. With non-differentiable sample paths, the weights are those of the Nadaraya-Watson estimator, that is, with the rule $0/0=0$, 
\begin{equation}
	W_m^{(i)}(t) = W_m^{(i)}(t;h) = K\left((T_m^{(i)} - t) / h \right)\left[\sum_{m=1}^{M_i} K\left((T_m^{(i)} - t) / h \right) \right]^{-1}.
\end{equation} 
The kernel $K$ is a symmetric density that, for simplicity, we can consider to be supported on $[-1,1]$. It is important to notice that some  values $\widehat X_t^{(i)}$ can be degenerate, in the sense that, for some points $t$, $W_m^{(i)}(t) = 0$ for all $1 \leq m \leq M_i$. This likely happens for some curves in the sparse independent design case. In the common design case, either all or none of the smoothed curves are degenerated at some points $t$. When $\widehat X_s^{(i)}$ or $\widehat X_t^{(i)}$ is degenerate,  the $i$-th curve will be dropped for covariance estimation at any point $(s,t)$. This issue of curve selection is not specific to our approach. It implicitly occurs with other estimators, see for example \cite{Li2010}, \cite{Zhang2016} and \cite{rubinP}. An adaptive bandwidth rule should regularize for the number of curves not used in estimation. By construction, our approach includes a penalization scheme, which automatically adapts to both sparse and dense regimes in a data-driven way. 

Let 
\begin{equation}
	w_i(t;h) = 
	\begin{cases}
		1 \qquad \text{if} \qquad \sum_{m=1}^{M_i} \mathbf{1}\{|T_m^{(i)} - t| \leq h\} \geq 1, \\
		0 \qquad \text{otherwise},
	\end{cases}
\end{equation}
indicate if there is at least one point along the curve $X^{(i)}$ to be selected in the estimation of  $X^{(i)}_t$. Let
\begin{equation}
	\mathcal{W}_N(t;h) = \sum_{i=1}^N w_i(t;h)\quad \text{and} \quad  \mathcal{W}_N(s,t;h) = \sum_{i=1}^Nw_i(t;h) w_i(s;h).
\end{equation}
 By construction, $w_i(t;h) = 0$ if and only if $W_m^{(i)}(t;h) = 0$ for all $1 \leq m \leq M_i$. 
With at hand the smoothed curves $\widehat X^{(i)}(h)$ defined as in \eqref{eq:smoothed_curves_x}, a natural covariance estimator would be
\begin{equation}\label{cov-function_x}
	 \widehat \Gamma_N(s,t;h) = \frac{1}{\mathcal{W}_N(s,t;h)}\sum_{i=1}^N w_i(s;h)w_i(t;h) \left\{\widehat X_t^{(i)}(h) - \widehat \mu_N(t;h)\right\}\left\{\widehat X_s^{(i)}(h) - \widehat \mu_N(s;h) \right\},\quad s,t\in\mathcal T,
\end{equation}
with  the mean estimator
\begin{equation}\label{mean-function}
	\widehat \mu_N(t;h) = \frac{1}{\mathcal{W}_N(t;h)}\sum_{i=1}^N w_i(t;h) \widehat X_t^{(i)}(h).
\end{equation}
The random positive integers $\mathcal{W}_N(t;h)$ and $\mathcal{W}_N(s,t;h)$ are thus the effective numbers of curves used in mean and covariance functions estimation, respectively.

It is well-known that a bias is usually induced on the diagonal set of the estimated covariance function. With a kernel $K$ supported on $[-1,1]$, the diagonal set is $|s-t|\leq 2h$. Since the eigen-elements are ultimately built by the eigen-decomposition of a covariance function, albeit with its own specific bandwidth, a correction is recommended to reduce the diagonal bias.  The nonparametric covariance estimator we propose is 
\begin{equation}\label{cov_est_corr}
	\widehat \Gamma (s,t;h) = \widehat \Gamma_N(s,t;h)  - 
	\widehat d_N(s,t;h),
\end{equation}
with the diagonal correction
\begin{equation}\label{diag_corr_eq}
\widehat d_N(s,t;h)=	
	\frac{\sigma(s) \sigma(t)}{\mathcal{W}_N(s,t;h)}\sum_{i=1}^N w_i(s;h)w_i(t;h) \sum_{m=1}^{M_i}W_m^{(i)}(s;h)W_m^{(i)}(t;h),
\end{equation}
where $\sigma^2(t)$ is the conditional variance defined in \eqref{data}. A nonparametric  estimator of $\sigma(t)$ is proposed in \eqref{eq:sigma-hat_x}, and further elaborated in the algorithm in Section \ref{sec:algorithm}.
By construction, $\widehat d_N(s,t;h)=0$ when $|s-t|>2h$, and thus $\widehat \Gamma   = \widehat \Gamma_N$ for $(s,t)$ outside the diagonal set.
 The `first smooth, then estimate' approach induces a bias of rate $o_{\PP}(1)$ on each curve. Our theoretical and empirical investigation show that the diagonal correction \eqref{diag_corr_eq} further improves the performance of the eigen-elements estimators. Our diagonal correction  is different from that in \cite{Golovkine2021}, since it corresponds to an integrated, instead of pointwise, risk.

The covariance estimator $\widehat \Gamma$ was formally defined in \eqref{cov_est_corr}  for a generic bandwidth $h$. The bandwidth used to compute the adaptive estimator of an eigen-element, will be  adapted for each eigen-element separately, and obtained by minimization of the corresponding  risk bound presented in Section \ref{sec:risk-sec_x}.

\subsection{Risk bounds}\label{sec:risk-sec_x}
 Let $\widehat \lambda_j$ and $\widehat \psi_j$ be the $j-$th eigenvalue and the associated eigenfunction, respectively, obtained from 
 our   covariance estimator \eqref{cov_est_corr}.  Let
\begin{equation}\label{eq:mon_o2}
	 \nu(t)  = \operatorname{Var}(X_t)   \qquad \text{and} \qquad c_2(s,t) = \operatorname{Var}(X_s X_t) ,\quad s,t\in\mathcal T.
\end{equation}
We  prove in Section \ref{sec:th_grnd} that, up to terms which are negligible or do not depend on the bandwidth, the eigenvalue risk $\mathcal{R}_N(\widehat \lambda_j ;h)$ can be bounded by twice
	\begin{multline}\label{eq:evalue-bound_x}
		\mathcal{B}_N(\widehat \lambda_j ;h) = 4 \int \nu(s)   \psi_{j}^2(s) ds \int L^2_t h^{2H_t}  \psi_{j}^2(t) \int |u|^{2H_t} K(u) du dt \\
		+ 
	 2	\iint \left\{\frac{ \sigma^2(t)\nu(s)}{\mathcal{N}_\Gamma(s|t;h)} + \frac{\sigma^2(s)\nu(t)}{\mathcal{N}_\Gamma(t|s;h)}\right\}   \psi_{j}^2(t)   \psi_{j}^2(s) dt ds  \\ 
		+ \iint c_2(s,t)\left\{\mathcal{W}_N^{-1}(s,t;h) - N^{-1}\right\}  \psi_{j}^2(t)   \psi_{j}^2(s) dt ds\\ =: \mathcal{B}_{1,N} (\widehat \lambda_j ;h) + \mathcal{B}_{2,N} (\widehat \lambda_j ;h) + \mathcal{B}_{3,N} (\widehat \lambda_j ;h),
	\end{multline}
where
\begin{equation}
	\mathcal{N}_\Gamma(t|s;h) = \left[\mathcal{W}_N^{-2}(s,t;h) \sum_{i=1}^N w_i(s;h)w_i(t;h)\max_{1 \leq m \leq M_i}|W_m^{(i)}(t;h)|\right]^{-1}.
\end{equation}
Meanwhile, the part of the eigenfunction risk $\mathcal{R}_N(\widehat \psi_j ;h)$ depending on $h$ can be bounded using 
\begin{multline}\label{eq:efunction-bound_x}
	 \mathcal{B}_N(\widehat \psi_j;h) = \sum_{k\in\mathcal K :k \neq j}\frac{2}{(  \lambda_{j} -   \lambda_{k})^2}\iint   \psi_{j}^2(t) 
	 \psi_{k}^2(s) \\ 
	 \qquad \times \left\{\nu(t) L^2_s h^{2H_s}\int |u|^{2H_s}K(u) du + \nu(s)L^2_t h^{2H_t}\int |u|^{2H_t}K(u) du  \right\} dt ds \\ 
+ \sum_{k\in\mathcal K :k \neq j} \frac{2} 
{(  \lambda_{j} -   \lambda_{k})^2} \left[\iint 
\psi_{k}^2(s)   \psi_{j}^2(t) \left\{ \frac{\sigma^2(s)\nu(t)}{\mathcal{N}_{\Gamma}(t|s;h)} + \frac{\sigma^2(t)\nu(s)}{\mathcal{N}_{\Gamma}(s|t;h)} \right\} dt ds \right] 
	\\
	+ \sum_{k\in\mathcal K :k \neq j}\frac{ 1 }{(  \lambda_{j} -   \lambda_{k})^2}\iint \psi_{k}^2(s)\psi_{j}^2(t) c_2(s,t)
	\left\{\mathcal{W}^{-1}_N(s,t;h) - N^{-1}\right\}    dt ds\\
	=: \mathcal{B}_{1,N} (\widehat \psi_j ;h) + \mathcal{B}_{2,N} (\widehat \psi_j ;h) + \mathcal{B}_{3,N} (\widehat \psi_j ;h).
\end{multline}
The summation indices in the last three sums belong to a finite set $\mathcal K$, for instance the integers from 1 to some threshold  $K_0$  to be set by the practitioner.  The bounds in \eqref{eq:evalue-bound_x} and \eqref{eq:efunction-bound_x} are derived from a generalized version of the eigen-elements representations given by \cite{Hall2006c} and \cite{HallNasab2009}, using the local regularity property \eqref{eq:def_lr1x}. Details are provided in the Appendix and the  Supplementary Material.

The terms $\mathcal{B}_{1,N} (\widehat \lambda_j ;h)$ and $\mathcal{B}_{1,N} (\widehat \psi_j ;h)$ are the squared bias terms which depend on the local regularity, while $\mathcal{B}_{2,N} (\widehat \lambda_j ;h)$ and $\mathcal{B}_{2,N} (\widehat \psi_j ;h)$ are the variance terms. The terms $\mathcal{B}_{3,N} (\widehat \lambda_j ;h)$ and $\mathcal{B}_{3,N} (\widehat \psi_j ;h)$ are the penalty terms which regularize for the curves which are dropped in the covariance estimation procedure.
It is likely that all curves are selected with large $h$, so that $\mathcal{W}_N(s,t;h) = N$ and the third term vanishes, getting us back to the classical bias-variance trade off. On contrary, low $h$ values lead to large $\mathcal{B}_{2,N}$ and $\mathcal{B}_{3,N}$ terms. In particular, in the common design case where all the  $M_i$ are equal and the set of $T_m^{(i)}$ is the same for all $i$, the bandwidth $h$ cannot go below the smallest distance between two consecutive observed time points $T_m^{(i)}$.
Our risk bounds realize an automatic balancing in the bandwidth selection process. 

Both squared bias terms $\mathcal{B}_{1,N} (\widehat \lambda_j ;h)$ and $\mathcal{B}_{1,N} (\widehat \psi_j ;h)$ feature the integrated regularity of $X$ represented by integrals of functions of $t$ involving the factor $h^{2H_t}$.  This governs the rate of convergence of the bias term, and a good estimator should adapt automatically to the unknown smoothness of the sample paths. We overcome this hurdle by explicitly estimating the regularity, and building estimators of eigen-elements that automatically take these smoothness estimates into account. It is important to notice that the rate of the integrals involving $h^{2H_t}$ are driven by the smallest values of $H_t$ when $h$ tends to zero.

The risk bounds presented above are tailored for the eigen-elements. Similar but different bounds can be derived, for example, for the estimation of the mean and covariance functions, and they also depend on the local regularity parameters $H$ and $L$. It turns out that the local regularity is a useful notion that allows to derive refined, exploitable risk bounds for nonparametric methods in FDA. A discussion on this issue can be found in Section \color{blue} V.4 \color{black} in the Supplement.

Practitioners should not be surprised to discover unaesthetic plots of the smoothed curves when constructed with our adaptive bandwidths. This is because although our bandwidths result in a good estimation of the eigen-elements, which is what they are optimised for, is not necessarily the optimal choice for curve smoothing purposes. This is because the risks are inherently different, which will result in different optimal bandwidths. More details and plots are provided in Section \color{blue} VI.6.1 \color{black} in the Supplement.

\subsection{Estimation of auxiliary quantities in the risk bounds}\label{sec:opt}

In addition to the functions $H$ and $L$ which characterize the local regularity of $X$, the risk bounds \eqref{eq:evalue-bound_x} and \eqref{eq:efunction-bound_x} contain several other quantities which  play the role of constants and can be easily estimated. We will see in Section \ref{sec:algorithm} that estimators for the moment functions $\nu(t)$ and $c_2(s,t)$ can be obtained as byproducts of the procedure for the estimation of $H$ and $L$.    In order to obtain preliminary estimates of eigen-elements  $ \lambda_j$ and  $\psi_j$ involved in the risk bounds, we  simply perform a preliminary data-driven eigen-elements estimation step where we minimize the risk bound \eqref{eq:evalue-bound_x} with the  $ \psi_j$'s replaced by  1. It is then easy to see from \eqref{eq:evalue-bound_x} and \eqref{eq:efunction-bound_x} that this simplification yields a same data-driven bandwidth for all eigenvalues and all eigenfunctions. We use it to compute   preliminary estimates  for  $\lambda_j$ and   $\psi_j$.


Finally, we propose a simple estimator for the conditional variance of the error terms, from which an estimate of $\sigma^2 (t)$ can be deduced. As presented in \eqref{data}, we here allow heteroscedastic error terms  
\begin{equation}\label{eq:hetero_e}
\varepsilon_m^{(i)}= \sigma(T_m^{(i)}) e_m^{(i)}, \qquad \forall 1\leq m\leq M_i, 1\leq i\leq N,
\end{equation} 
where   the $e_m^{(i)}$'s are a random sample of a standardized variable $e$.
Whenever the observed time points $T_m^{(i)}$ admit a density $f_T$ which is bounded away from zero on $\mathcal T$, a simple  estimator of  $\sigma^2(t)$ is
\begin{equation}\label{eq:sigma-hat_x}
   	\widehat \sigma^2(t;b) = \frac{1}{2\mathcal I_N(t;b)}\sum_{i=1}^N \! \left(Y_{m(t,i,1)}^{(i)} - Y_{m(t,i,2)}^{(i)}\right)^{\!2} \mathcal I^{(i)}(t;b)  \quad \text{with} \quad \mathcal I^{(i)}(t;b) = \mathbf{1}\left\{ \left|T_{m(t,i,2)}^{(i)} \!-t\right| \leq b \right\},  
\end{equation}
where $\mathcal I_N(t;b)=\sum_{i=1}^N \mathcal I^{(i)}(t;b)  $,  $b>0$. For each $i$, $(Y_{m(t,i,1)}^{(i)},T_{m(t,i,1)}^{(i)})$ and  $(Y_{m(t,i,2)}^{(i)},T_{m(t,i,2)}^{(i)})$ are the two pairs with the observed time points closest and the second closest to $t$, respectively. The estimator $	\widehat \sigma^2(t;b)$ is inspired by \cite{stad87} and adapted for FDA.
Our estimator adjusts for possibly large spacings between the observed time points, which for instance is more relevant to the sparse regime. It suffices to average over the values $i$ for which  $\big|T_{m(t,i,1)}^{(i)}\!-T_{m(t,i,2)}^{(i)}\big|$ is sufficiently small,    according to the value of $b$, which decreases to zero with the sample size. The rate of decrease of $b$ has little impact when $\widehat \sigma^2(t;b)$ is plugged into the risk bounds \eqref{eq:evalue-bound_x} and \eqref{eq:efunction-bound_x}, as only consistency is required at that level. However, the decrease of $b$ should be suitably set when $\widehat \sigma^2(t;b)$ is used in the final diagonal bias correction \eqref{diag_corr_eq}. Details on the data-driven rules to choose $b$ in both situations are provided in the next section.



\subsection{The algorithm}\label{sec:algorithm}

We now summarize our adaptive FPCA methodology    which is  based on the minimization of the  risk bounds 
\eqref{eq:evalue-bound_x} and \eqref{eq:efunction-bound_x} with respect to $h$ in a bandwidth range $\mathcal H_N$. Our algorithm is fast and the tuning parameters are fixed in a data-driven way. Since the risk bounds depend on the unknown eigenvalues  $ \lambda_j$ and eigenfunctions $ \psi_j$, to make our procedure fully self-contained, we propose to run it twice, as explained at the end of this section.  

\bigskip


\hrule
\textbf{Adaptive FPCA Algorithm }
\hrule

\begin{enumerate}[Step 1:]
	\item[\!\!\textbf{Input} ]\label{step0} Proxies of  $ \lambda_j$ and $ \psi_j$; default values for the  bandwidth $b$;   $\zeta\in(0,1/2)$; integer threshold $K_0$;  bandwidth range $\mathcal H_N$.  
	
\item\label{step1} 	\textbf{Presmoothing.} Presmooth each curve $i$, for example using the Nadaraya-Watson estimator with a simple  bandwidth rule. Denote by $\widetilde X_t^{(i)}$ the presmoothing estimator of $X^{(i)}$ at point $t$.
	
\item\label{step2} \textbf{Estimation of $H$ and $L$.} 
For any $t$, choose suitable $t_1,t_2,t_3\in\mathcal T$,
such that $|t_1-t_3 | = 2|t_1-t_2|$ and $t$ belongs to the interval defined by $t_1$ and $t_3$.  Then build estimates 
\begin{equation}\label{eq:hat-h_x}
\widehat H_t = \frac{\log\!\big( \widehat\theta(t_1,t_3)\big)\! - \log\! \big(\widehat\theta(t_1,t_2)\big)}{2\log (2)} , 
\quad 
 \widehat L_t ^2= \frac{\widehat\theta(t_1,t_3)}{|t_1\!-t_3|^{2  \widehat H_t }} \quad \text{ with } \quad 	\widehat\theta(u,v) = \frac1{N} \sum_{i=1}^{N}
 \left\{\!
 \widetilde{X}^{(i)}_u \!- \widetilde{X}^{(i)}_v
 \!\right\}^2\!.
\end{equation}

\item\label{step3} \textbf{Estimation of the moment functions.} Build the estimates
\begin{equation}
\widehat \nu(t) = \frac{1}{N}\sum_{i=1}^N \left\{ \! \widetilde X_t^{(i)}\!\right\}^2 -
\left\{\!\frac{1}{N}\sum_{i=1}^N   \widetilde X_t^{(i)}\!\right\}^{\!2}, \quad\widehat c_2(s,t)	
= \frac{1}{N}\sum_{i=1}^N \! \left\{ \!\widetilde X_s^{(i)} \widetilde X_t^{(i)}\! \right\}^2 
- \left\{\!\frac{1}{N}\sum_{i=1}^N  \widetilde X_s^{(i)} \widetilde X_t^{(i)}\!\right\}^{\!2}
,
\end{equation}
of $\nu(t) = \operatorname{Var}(X_t)$ and $c_2(s,t) = \operatorname{Var}(X_sX_t)$, respectively. 

\item\label{step3b} \textbf{Estimation of the conditional variance.} Apply \eqref{eq:sigma-hat_x} and build 
the estimate  $\widehat \sigma^2(t;b)$ of the errors' conditional variance $\sigma^2(t)$.

\item\label{step4} \textbf{Minimize the risk bounds with respect to $h$.} Let $\widehat {\mathcal{B}}_N(\widehat\lambda_j ;h) $ and $\widehat {\mathcal{B}}_{N} (\widehat\psi_j ;h)$
be the functions of $h$ obtained by replacing in  the expressions of $\mathcal{B}_N(\widehat \lambda_j ;h) $ and $\mathcal{B}_{N} (\widehat \psi_j ;h)$ the unknown quantities with the estimates from Step \ref{step3}  and Step \ref{step3b}. Numerically approximate the integrals in \eqref{eq:evalue-bound_x} and \eqref{eq:efunction-bound_x} and compute $\widehat {\mathcal{B}}_N(\widehat\lambda_j ;h) $ and $\widehat {\mathcal{B}}_{N} (\widehat \psi_j ;h)$ for $h$ on a grid in a range of  bandwidths $\mathcal{H}_N$. For the computation of $\widehat {\mathcal{B}}_{N} (\widehat\psi_j ;h)$, take $\mathcal K = \{1,\ldots,K_0\}$ in \eqref{eq:efunction-bound_x}. Select the bandwidths $\widehat h^*(\lambda_j)$ and $\widehat h^*(\psi_j)$ that minimize $\widehat {\mathcal{B}}_N(\widehat\lambda_j ;h) $ and $\widehat {\mathcal{B}}_{N} (\widehat\psi_j ;h)$, respectively.  

\item\label{sze:tep5} \textbf{Adaptive covariance function estimation and eigen-decomposition.} For each $j\geq 1$, compute the covariance estimates $\widehat \Gamma(s,t;\widehat h^*(\lambda_j))$ and $\widehat \Gamma(s,t;\widehat h^*(\psi_j))$ using definition \eqref{cov_est_corr}. For the diagonal corrections \eqref{diag_corr_eq}, use $\widehat \sigma^2(\cdot;b^*)$ from  \eqref{eq:sigma-hat_x} instead of $\sigma^2(\cdot)$, with $b^*(\lambda_j)=\{\widehat h^*(\lambda_j) \}^{1-\zeta}$ and  $b^*(\psi_j)=\{\widehat h^*(\psi_j) \}^{1-\zeta}$, respectively. 
 Build the adaptive estimates $\widehat \lambda_j$ and $\widehat \psi_j$ with the covariance estimates  $\widehat \Gamma(s,t;\widehat h^*(\lambda_j))$ and $\widehat \Gamma(s,t;\widehat h^*(\psi_j))$, respectively.

%
\end{enumerate}
\hrule

\vspace{.7cm}

For presmoothing in Step \ref{step1} we propose a simple  least-squares cross-validation (LS-CV) approach. We use LS-CV on a small subset of curves, for instance 20 curves, and take the median of the 20 bandwidths selected. 
With this bandwidth we smooth separately each curve $i$ in the sample at the observed points $T_m^{(i)}$, and between the $T_m^{(i)}$'s we simply  use linear interpolation. Details on the choice of $t_1,t_3$ in Step \ref{step2} are provided in Theorem \ref{thm:estimation-alpha_L} below. A rich toolkit for practitioners to perform eigenanalysis in Step \ref{sze:tep5} is available, see for instance Chapter 8.4 of \cite{Ramsay2005}. 


For a self-contained procedure, we first run the algorithm with a \emph{ad-hoc} proxy for all $ \psi_j$ considered, that is the constant function equal to 1. The $ \lambda_j$'s then 
no longer matter in the optimization of the feasible bounds  $\widehat {\mathcal{B}}_N(\psi_j ;h)$. In this simplified first run, since $b$ only serves to calculate constants, with independent design we propose a default choice  $b=\operatorname{length}(\mathcal T)/10$. In the common design case, we simply take $b$ equal to the largest spacing between consecutive observed points $T_m^{(i)}$ and $T_{m+1}^{(i)}$. 
The simplified first run yields preliminary estimates for $\lambda_j$ and $ \psi_j$ obtained in Step \ref{sze:tep5} from a single adaptive, data-driven bandwidth, say, $\widehat h^*$. For the diagonal correction, we use the variance estimator with $b=\{\widehat h^* \}^{1-\zeta}$, for some $\zeta >0$. We propose the default value $\zeta = 0.1$.
Next, in both independent and common design cases, we run again the Steps \ref{step4} to \ref{sze:tep5} of the algorithm with the preliminary estimates of $ \lambda_j$ and $ \psi_j$. We then have an optimal bandwidth $\widehat h ^*$ and an associated $b^*$ for each eigen-element separately. The rationale for the choice of $b^*$ in Step \ref{sze:tep5} is provided by Lemmas A.\ref{lem:cvp_sigma} and A.\ref{repair_diag}  in the Appendix.

Using different bandwidths for different eigenfunctions, which is expected to improve accuracy, may break their orthogonality. This inconvenience can be corrected by a simple Gram-Schmidt procedure. However, our extensive simulation study shows that the scalar product between our eigenfunction estimates is usually very close to zero. Empirical evidence is provided in Section \color{blue} VI.6.2 \color{black} of the Supplement.

We learn from  Lemma \ref{lemma_Ih} in the Appendix, that numerical approximation of 
the squared  bias terms $\mathcal{B}_{1,N} (\widehat \lambda_j ;h)$ and $\mathcal{B}_{1,N} (\widehat \psi_j ;h)$ could artificially increase the squared bias by a $\log(1/h)$ factor. Then the Algorithm would yield an optimal bandwidth diminished by a logarithmic factor. Based on these considerations and our extensive simulation study, an upward correction of the values $\widehat h^*(\lambda_j)$ and $\widehat h^*(\psi_j)$ obtained by minimization of $\widehat {\mathcal{B}}_{N} (\widehat \lambda_j ;h)$ and $\widehat {\mathcal{B}}_{N} (\widehat \psi_j ;h)$ may be applied for improved practical results. The final bandwidths will then be given by $\widehat  h _{WPK}^*(\lambda _{j})   = 
\log (1/\widehat h^*(\lambda_j)) 
 \times  \widehat h^*(\lambda_j)$ and $ \widehat  h _{WPK}^*(\psi _{j}) = \log (1/\widehat h^*(\psi_j)) \times  \widehat h^*(\psi_j)$.
This logarithmic correction factor becomes closer to 1 when the bandwidths $\widehat h^*$ are larger.


\section{Theoretical properties}\label{sec:th_grnd}

Our approach adapts to the regularity of the process $X$, a notion informally introduced in \eqref{eq:def_lr1x}. The idea of local regularity estimators \eqref{eq:hat-h_x} was introduced by \cite{Golovkine2022}. For the purpose of FPCA, we reconsider their construction and provide new, non-asymptotic exponential bounds for the uniform concentration of the regularity estimators $\widehat H$ and $\widehat L$, a result of independent interest. However, for the sake of readability and given our focus on the  new approach for FPCA, we postpone the formal definition of the local regularity and the uniform concentration bounds of $\widehat H$ and $\widehat L$ to the Appendix. See 
 Sections \ref{sec:def_loc_reg_x} and \ref{ssec:est_x}.

In the following, we state the formal results on the risk bounds for the risks defined in \eqref{risk_def1}. Next, we provide the rate of the bandwidths minimizing the risk bounds, from which we derive the rates of convergence for our  adaptive 
estimators of the eigen-elements. Finally, we show that using the feasible risk bounds $\widehat {\mathcal{B}}_N(\widehat\lambda_j ;h) $ and $\widehat {\mathcal{B}}_{N} (\widehat\psi_j ;h)$, where the unknown quantities are replaced by estimates, instead of  $\mathcal{B}_N(\widehat \lambda_j ;h) $ and $\mathcal{B}_{N} (\widehat \psi_j ;h)$,  does not alter the rates of convergence. The assumptions used for the theoretical results are provided in the Appendix \ref{sec_ap:ass}. Recall that we consider $\lambda_j$  decreasingly ordered.

\subsection{Risk bounds and rates of convergence for eigen-elements estimators}\label{sec:risk-sec}

Our first result shows that, modulo a constant and negligible terms not depending on $h$, the quadratic risks  $\mathcal{R}_N(\widehat \lambda_j ;h)$ and $\mathcal{R}_N(\widehat \psi_j ;h)$ can be bounded by the quantities  defined in \eqref{eq:evalue-bound_x} and \eqref{eq:efunction-bound_x}, respectively. 

\begin{theorem}\label{thm:thm-1}
Let Assumptions \ref{ass_data} and \ref{ass_ad_smo} in Appendix \ref{sec_ap:ass} hold true, and $\mathcal H_N$ be a bandwidth range as in Assumption \ref{ass_ad_smo}.
For $j\geq 2$, assume 
$$
\min(\lambda_j-\lambda_{j+1}, \lambda_{j-1} - \lambda_j ) > 0,
$$ 
and $\lambda_1-\lambda_{2}>0$ when $j=1$. Then, uniformly over $\mathcal H_N$,
    \begin{equation}\label{eq:evalue-bound}
        \mathcal{R}_N(\widehat \lambda_j ;h) \leq 2 \mathcal{B}_N(\widehat \lambda_j;h)\{1+ o_{\mathbb P}(1)\} + O_{\mathbb P} (N^{-1}), \qquad \text{as } N,\mathfrak m \rightarrow \infty. 
    \end{equation}
Moreover, given a constant $C>0$, an integer $K_0$ depending on $C$ exists such that, for $\mathcal{B}_N(\widehat \psi_j;h)$ defined as in \eqref{eq:efunction-bound_x} with $\mathcal K = \{1,\ldots,K_0\}$, we have
\begin{equation}\label{eq:efunction-bound}
		\mathcal{R}_N( \widehat \psi_j;h)\leq  \{2 + C\}\mathcal{B}_N(\widehat \psi_j;h)\{1+ o_{\mathbb P}(1)\}  + O_{\mathbb P} (N^{-1}), \qquad  \text{as } N,\mathfrak m \rightarrow \infty. 
\end{equation}
\end{theorem}

\vspace{-0.3cm}

The  $O_{\mathbb P} (N^{-1})$ terms in Theorem \ref{thm:thm-1} are given by the risk of the empirical eigenvalues and eigenfunctions, respectively; see \eqref{ry_dec} and \eqref{ry_dec2}. We next study the rates of the bandwidths that minimize the risk bounds $\mathcal{B}_N(\widehat \lambda_j ;h)$ and $\mathcal{B}_N(\widehat \psi_j;h)$. These drive the rates of convergence of the quadratic risks $\mathcal{R}_N(\widehat \lambda_j ;h) $ and $	\mathcal{R}_N( \widehat \psi_j;h)$, respectively. The rates of $\mathcal{B}_N(\widehat \lambda_j ;h)$ and $\mathcal{B}_N(\widehat \psi_j;h)$, and the corresponding minimum $h$, 
are determined by
$$
\underline{ H} = \min_{t\in\mathcal T} H_t.
$$
While in practice the numerical optimization will select the bandwidth without particular difficulty, in general the exact theoretical rate of the bandwidth can only be determined up to a logarithmic factor; see Lemma A.\ref{lemma_Ih} in the Appendix. The reason is the exact rate of an integral involving the factor $h^{2H_t}$, which  depends on the behavior of $H_t$ in the neighborhood of its minimum. If for instance, $H$ is constant equal to $\underline{ H} $ on a non-degenerate interval, the integrals have the rate given by $h^{2\underline{ H}}$. If $\underline{ H}$ is attained at isolated  points, the rate of the integrals depend on the local behavior of $H$ in the neighborhood of these points. 
We therefore present a range of rates for the bandwidth and an upper bound for the rates of the risks, under the mild condition of Lipschitz continuity for the function $H$. 

Let us first consider the case of independent design and define
\begin{equation}\label{low_rate}
	\underline r(N,\mathfrak m) = \max\left\{\!\left(\!\frac{1}{N\mathfrak{m}^2} \! \right)^{\frac{1}{2 \underline{ H} + 2}},  \left(\!\frac{1}{N\mathfrak{m}}\!\right)^{\frac{1}{2\underline{ H} + 1}}\! \right\}, \qquad 	\overline r(N,\mathfrak m) =\max\left\{\!\left(\frac{\log(N\mathfrak{m})}{N\mathfrak{m}^2} \! \right)^{\frac{1}{2 \underline{  H} + 2}}, \left(\!\frac{\log(N\mathfrak{m})}{N\mathfrak{m}}\!\right)^{\frac{1}{2\underline{  H} + 1}}\!\right\}.
\end{equation}
In the following, the symbol $\lesssim$ means that the left side is bounded by a constant times the right side. Moreover, for a  sequence of strictly positive random variables $U_n$ and a sequence of  numbers $ b_n>0$, the notation  $ U_n\lesssim_{\PP}b_n$ means $U_n = O_{\PP}(b_n)$, while $b_n\lesssim_{\PP} U_n$ means $1/U_n = O_{\PP}(1/b_n)$. 

\begin{corollary}\label{corr_rates_est}
Assume  the conditions of Theorem \ref{thm:thm-1} hold true. 
Let $h^*(\widehat \lambda_j)$ and $  h^*(\widehat \psi_j) $ be the bandwidths obtained by minimization over $\mathcal H_N$ of 
$\mathcal{B}_N(\widehat \lambda_j;h)$ and $\mathcal{B}_N(\widehat \psi_j;h)$, respectively. Then, in the case of independent design, 
\begin{equation}\label{rates_h}
 \underline r(N,\mathfrak m) \lesssim_{\PP}   h^*(\widehat \lambda_j),  h^*(\widehat \psi_j) 
 \lesssim_{\PP} \overline r(N,\mathfrak m).
\end{equation}
Moreover, if $\widehat \lambda_j$ and $\widehat \psi_j$ are obtained with the bandwidths $h^*(\widehat \lambda_j)$ and $  h^*(\widehat \psi_j) $, respectively, then 
\begin{equation}
\EE\left[\left\{\widehat \lambda_j - \lambda_j \right\}^2\right] + \EE\left\|\widehat \psi_j - \psi_j \right\|_2^2 \lesssim
\overline r(N,\mathfrak m)^{2\underline H}
+ N^{-1}  .
\end{equation}
\end{corollary}

\smallskip

For simplicity, in the case of common design, we impose the following mild assumption on the spacings between consecutive observed time points: with $T_m^{(i)}$ increasingly ordered, a constant $C_d>0$ exists such that 
\begin{equation}\label{sp_cd}
 C^{-1}_d \mathfrak m^{-1} \leq \left |T_{m+1}^{(i)} -T_m^{(i)} \right|	\leq C_d \mathfrak m^{-1},  \qquad \forall m \in \{1,\ldots, \mathfrak m - 1\}.
\end{equation}

\smallskip

\begin{corollary}\label{corr_rates_est_cd}
	Assume  the conditions of Theorem \ref{thm:thm-1} and \eqref{sp_cd} hold true.  Let $h^*(\widehat \lambda_j)$,  $  h^*(\widehat \psi_j) $, $\widehat \lambda_j$ and $\widehat \psi_j$ be defined as in Corollary \ref{corr_rates_est}.  Then, in the case of common design, 
	\begin{equation}\label{rates_h}
\max\left\{  \mathfrak m^{-1} , (N\mathfrak{m})^{- \frac{1}{2\underline{ H} + 1}} \right\}
 \lesssim_{\PP}   h^*(\widehat \lambda_j),  h^*(\widehat \psi_j) 
\lesssim_{\PP}   \max\left\{ \log(\mathfrak m)/ \mathfrak m  , \{\log(N\mathfrak{m})/(N\mathfrak{m})\}^{ \frac{1}{2\underline{ H} + 1}} \right\}.
	\end{equation}
	Moreover, 
	\begin{equation}
		\EE\left[\left\{\widehat \lambda_j - \lambda_j \right\}^2\right] + \EE\left\|\widehat \psi_j - \psi_j \right\|_2^2 \lesssim 
		\{\log( \mathfrak m)/\mathfrak m\}^{2\underline H}
		+ N^{-1}.
	\end{equation}
\end{corollary}

\smallskip

There are significant differences in the rates for the common design case. These are caused by the penalty terms $\mathcal{B}_{3,N} (\widehat \lambda_j ;h)$ and $\mathcal{B}_{3,N} (\widehat \psi_j ;h)$ in \eqref{eq:evalue-bound_x} and \eqref{eq:efunction-bound_x}  which prevent the bandwidths to be smaller than  half of the spacings between consecutive observed time points. This aspect is automatically taken into account by our risk bounds which can be used with both  independent or common design.

Finally,  it remains to show that, up to constants and uniformly with respect to the bandwidth, the difference between  the estimated bounds  $\widehat {\mathcal{B}}_N(\widehat\lambda_j ;h) $ and $\widehat {\mathcal{B}}_{N} (\widehat\psi_j ;h)$ from Step \ref{step4} of the Algorithm in Section \ref{sec:algorithm}, and the infeasible  ones in \eqref{eq:evalue-bound_x} and \eqref{eq:efunction-bound_x}, is negligible. The following result applies to both first and second runs of the Algorithm.

\begin{theorem}\label{thm:thm-1-bis}
	Let   Assumptions \ref{ass_data}, \ref{ass_pr_sm} and \ref{ass_ad_smo} in Appendix \ref{sec_ap:ass} hold true. 
	Then, uniformly over $\mathcal H_N$,  
	\begin{equation}\label{eq:evalue-bound_2}
\widehat{\mathcal{B}}_N(\widehat \lambda_j;h) ={\mathcal{B}}_N(\widehat \lambda_j ;h) \{1+o_{\mathbb P}(1)\}   \quad \text{ and } \quad 
\widehat{\mathcal{B}}_N(\widehat \psi_j;h) =	 {\mathcal{B}}_N(\widehat \psi_j ;h) \{1+o_{\mathbb P}(1)\}, \quad \text{as } N,\mathfrak m \rightarrow \infty. 
	\end{equation}
\end{theorem}

\smallskip

Theorem \ref{thm:thm-1-bis} guarantees that Corollaries \ref{corr_rates_est} and \ref{corr_rates_est_cd} remain true with $h^*(\widehat \lambda_j)$ and $  h^*(\widehat \psi_j)$ replaced by the feasible versions $\widehat h^*(\widehat \lambda_j)$ and $  \widehat h^*(\widehat \psi_j)$, and the eigen-elements $\widehat \lambda_j$ and $\widehat \psi_j$ obtained with these bandwidths.


\section{Numerical properties}\label{sec:simu-sec}

We below provide  the results of an extensive  simulation study. The simulated data mimic the characteristics of a real data set of power consumption curves described in Section \ref{subsec:dgp_1}. The results are obtained using the package \texttt{FDAdapt} dedicated to our adaptive FPCA. The new, adaptive approach performs well when compared to existing FPCA approaches. 

Simulated functional data are generated according to property \eqref{eq:def_lr1x}.
For this purpose, we build in Section \ref{sec:sim_design} a wide class of Gaussian processes which satisfy \eqref{eq:def_lr1x} with given $H$ and $L$. This class  is interesting \emph{per se}, as it offers an easy to implement and effective simulation setup that allows the data generating process (DGP) to inherit the characteristics of real data.

\subsection{A general purpose simulator}\label{sec:sim_design}

The class of processes satisfying \eqref{eq:def_lr1x} is general. 
Examples include, but are not limited to stationary or stationary increment processes. See \cite{Golovkine2022} for examples.   Here, we consider the example of multifractional Brownian motion (MfBm) processes. See, \emph{e.g.}, \citet{balanca2015} and the references therein for the formal definition. 
The MfBm, say $W=(W(t))_{t\geq 0}$, with Hurst index function, say $t\mapsto H_t \in(0,1)$, is a centered Gaussian process with covariance function 
\begin{equation}
	C(s,t) = \EE\left[W(s)W(t)\right] =  D(H_s,H_t )\left[ s^{H_s+H_t} +  t^{H_s+H_t} - |t-s|^{H_s+H_t}\right] , \qquad s, t\geq 0,
\end{equation}
where
\begin{equation}\label{eq_D}
	D(x,y )=\frac{\sqrt{\Gamma (2x+1)\Gamma (2y+1)\sin(\pi x)\sin(\pi y)}} {2\Gamma (x+y+1)\sin(\pi(x+y)/2)} , \qquad D(x,x) = 1/2, \qquad x,y >0.
\end{equation}
  We show in Lemma SM.\color{blue}1 \color{black} in the Supplementary Material that $W$ satisfies \eqref{eq:def_lr1x} with  the Hurst index as exponent and $L\equiv 1$.   The  fractional Brownian motion is an example of MfBm with constant Hurst index.

In order to accommodate more complex covariance functions that may be encountered in practice, in particular to allow more general functions $L$ and variance functions,  we consider a deterministic time deformation defined by a map $A:\mathbb R_+ \rightarrow \mathbb R_+$. This map is assumed to be continuously differentiable, with strictly positive derivative. Let $A^{-1}$ be the inverse of $A$, and 
\begin{equation}
	H_{A,u} = H_{A^{-1}(u)}.
\end{equation}
Given the function $H$ and the time deformation $A$, the processes $X$ we consider in our simulations are defined as
	\begin{equation}\label{eq:X_md}
		X(t) = \mu(t) + \tau(t)W_A (A(t)), \qquad t\geq 0,
	\end{equation}
	where  $(W_{A}(u))_{u\geq 0}$ is a MfBm with Hurst index function $H_{A,u}$,  $\mu$ is a mean function, and $\tau$ is a scaling function. The covariance function of the Gaussian process $X$ is 
\begin{equation}\label{eq:cov_mfbm_c}
	C_A(s,t) 
	= \tau(s)\tau(t) D(H_s,H_t )\left[ A(s)^{H_s+H_t} +  A(t)^{H_s+H_t} - |A(t)-A(s)|^{H_s+H_t}\right] ,
\end{equation}	
and allows for flexible patterns matching real data characteristics. It suffices to suitably define the deformation $A$ and the scaling $\tau$. In our simulations, we consider two situations. First, we consider $A$ and $\tau$ such that $X$ defined in \eqref{eq:X_md} satisfies \eqref{eq:def_lr1x} with given functions $H$ and $L$. In this case, we take  
\begin{equation}\label{eq:A-construct_0}
A(t) = A(0)+ \int_{0}^t L_s^{1/H_{s}} ds, \quad A(0)\geq 0,\qquad \text{and} \quad \tau(t) =1, \;t\in\mathcal T,
\end{equation}
and we have  $\operatorname{Var}(X(t))=  A(t)^{2H_t}$. In the second case, we want $X$ to match functions  $H$, $L$ and a given variance function $\nu(t)=\operatorname{Var}(X(t)) $. For this, we  consider
\begin{equation}\label{eq:A-construct}
	A(t) = A(0)\exp\left(\int_{0}^t  \{\nu (s)^{-1} L_s\} ^{1/H_{s}} ds\right), \quad 
	\;A(0)>0, \qquad \text{and} \quad \tau(t) = \nu(t) A(t)^{-H_{t}} , \;t\in\mathcal T.
\end{equation}
Then  $X$ defined in \eqref{eq:X_md} still  satisfies \eqref{eq:def_lr1x}, and  $\operatorname{Var}(X(t))=  \nu(t)$. 
  The justification of these statements is provided in Section \color{blue} I \color{black}
of the Supplementary Material.


\subsection{Data generating process}\label{subsec:dgp_1}

We construct a simulation DGP with characteristics close to those of  a data set containing household electricity consumption, which can be downloaded at \path{https://archive.ics.uci.edu/dataset/235/individual+household+electric+power+consumption}. Different electrical quantities are measured, with 1 minute sampling rates over a period of 4 years, resulting in almost 2 million data points. 

\begin{figure}[htp]
\centering
\includegraphics[width=.33\textwidth]{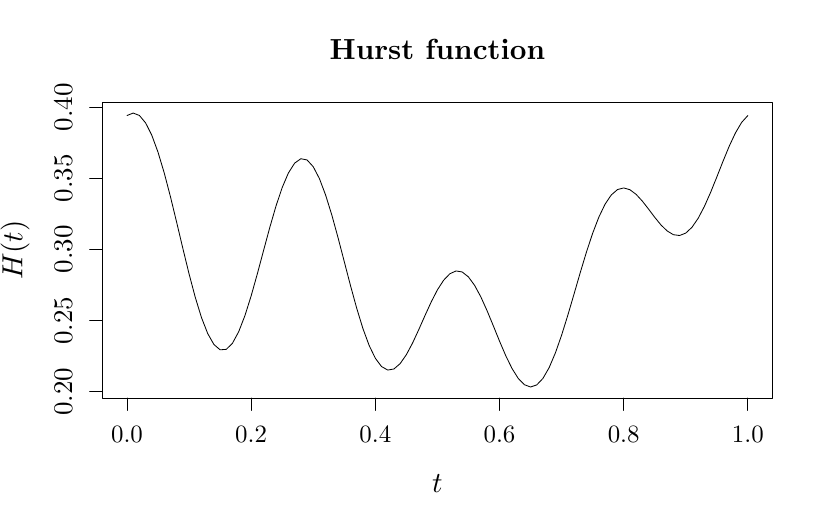}\hfill
\hspace{-0.4cm}\includegraphics[width=.33\textwidth]{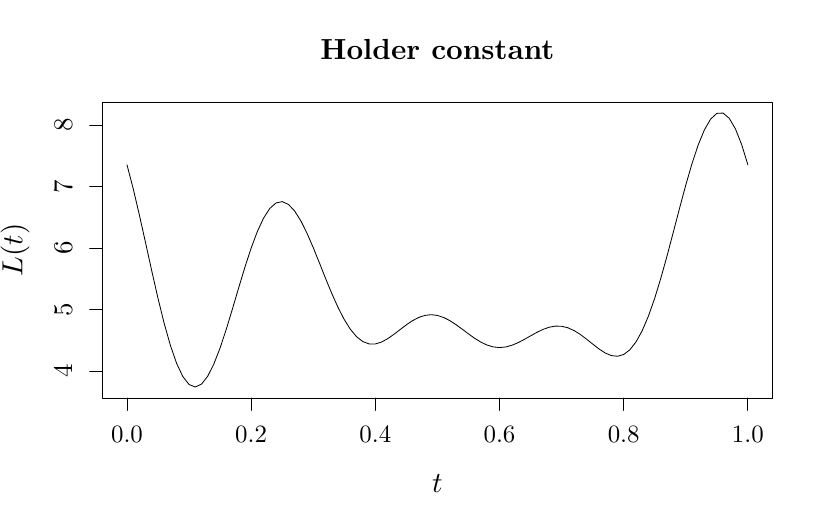}\hfill
\hspace{-0.4cm} \includegraphics[width=.34\textwidth]{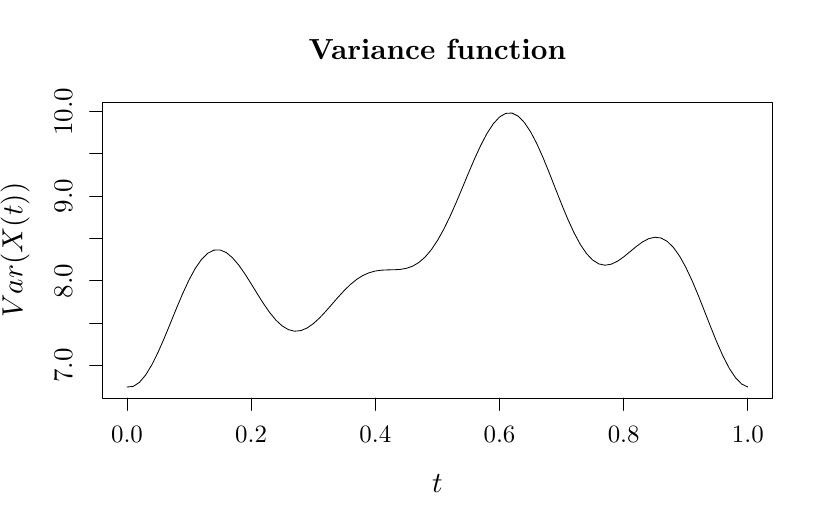}
\vspace{-.2cm}
\caption{\small Simulation DGP:  regularity parameters $H$ (left), $L$ (middle) and the variance function $\nu$ of $X$ (right)}
\label{fig:regularity_plot}

\end{figure}

To restrict ourselves to an univariate analysis, we focus on  the daily voltage variable, considering days without missing values. This data subset contains 1351 voltage curves observed on an equally spaced grid of 1440 sampling points which are re-scaled on $[0,1]$. We estimate the local regularity parameters $H_t$ and $L_t$ according to Step \ref{step2} of the Algorithm, with simple linear interpolation presmoothing. See also the procedure described in the Section \ref{ssec:est_x}. The variance function $\nu(t) = \operatorname{Var}(X_t)$ is similarly estimated by interpolation according to Step \ref{step3}. These three quantities were estimated on a grid containing 40 equally spaced evaluation points in the unit interval.
We then smooth the estimated parameters with 9 Fourier basis functions.
The constructed functions can be seen in Figure \ref{fig:regularity_plot}. Finally, we also define a mean function $\mu(t)$  as the empirical mean of the densely sampled electricity curves, shifted downward by 240, and smoothed by running a LASSO regression on linear combinations of sine and cosine functions using the \texttt{glmnet} package. 
Additional details on the real and simulated data are provided in Section \color{blue} VI \color{black} in the Supplement.

\begin{figure}[ht]
	\vspace{-.25cm}
\centering
\vspace{-.5cm}
\includegraphics[height = 0.23\textheight,width=.3\textwidth]{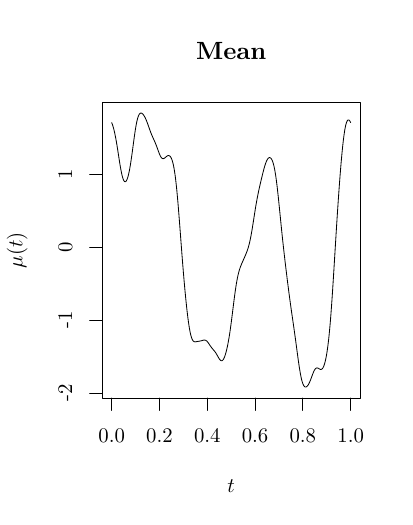}
\includegraphics[height = 0.23\textheight,width=.3\textwidth]{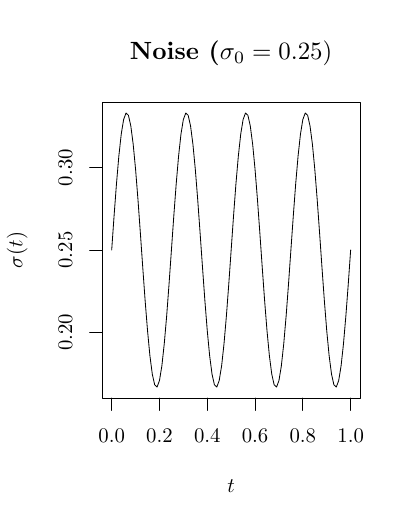}\hfill
\includegraphics[height = 0.23\textheight,width=.3\textwidth]{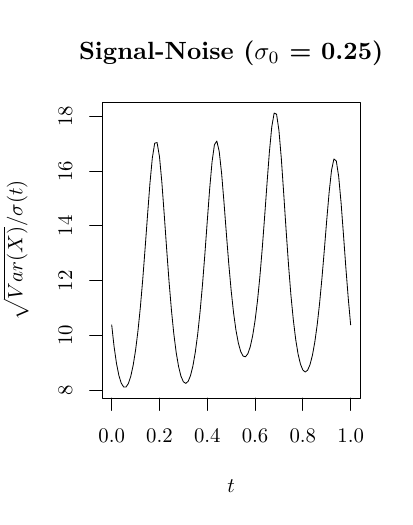}\hfill
\vspace{-.5cm}
\caption{\small  Simulation DGP:  mean  $\mu$ (left), conditional standard deviation  $\sigma$ (middle), Signal-to-Noise Ratio (right) }
\label{fig:noise_plot}
\end{figure}

A time deformation $A(t)$ and a scaling function $\tau(t)$ are built using \eqref{eq:A-construct} and the three constructed functions $H$, $L$ and $\nu$. The covariance function \eqref{eq:cov_mfbm_c} and the mean function $\mu(t)$ are then used to simulate the sample paths. More precisely, for each $1\leq i\leq N$, after drawing $M_i$ and the  $T_m^{(i)}$'s, we use that covariance function and the mean vector $(\mu(T_1^{(i)}),\ldots, \mu(T_{M_i}^{(i)}))$ to generate the Gaussian random vectors $(X^{(i)}(T_1^{(i)}),\ldots, X^{(i)}(T_{M_i}^{(i)}))$, and ultimately the values $Y_1^{(i)},\ldots,Y_{M_i}^{(i)}$ using  \eqref{data}, by adding a heteroscedastic noise with conditional standard deviation $\sigma(t) = \sigma_0 (1+\sin(8\pi t)/3)$. Two levels of noise are considered corresponding to $\sigma_0=0.25$ and $\sigma_0 = 1$, respectively. In the higher noise case, the signal-to-noise is thus divided by 4. Graphs of the mean $\mu(t)$, the noise standard deviation $\sigma(t)$ and of the signal to noise ratio in the case of lower noise are shown in Figure \ref{fig:noise_plot}.

\begin{table}
\small 	\centering\caption{\small  Simulation DGP: eigenvalues \& their cumulative explained variance (CEV) in percentage}
	\begin{tabular}{l|c|c|c|c|c|c|c|c|c}
	 & $\lambda_1$ & $\lambda_2$  &  $\lambda_3$ & $\lambda_4$ & $\lambda_5$ & $\lambda_6$ & $\lambda_7$ & $\lambda_8$ & $\lambda_9$ \\ \hline
		Eigenvalue & 7.354 & 0.296 & 0.152 & 0.074 & 0.039 & 0.035 & 0.023 &  0.017 & 0.016 \\ \hline
		CEV (\%)& 89.7 & 93.3 & 95.2 & 96.1& 96.6 & 97 & 97.3 & 97.5 & 97.7
	\end{tabular}\label{evalues-table}
\end{table}

\begin{figure}[!ht]
\centering
\vspace{-.25cm}
\hspace{-1cm} \includegraphics[height = 0.26\textheight,width=.49\textwidth]{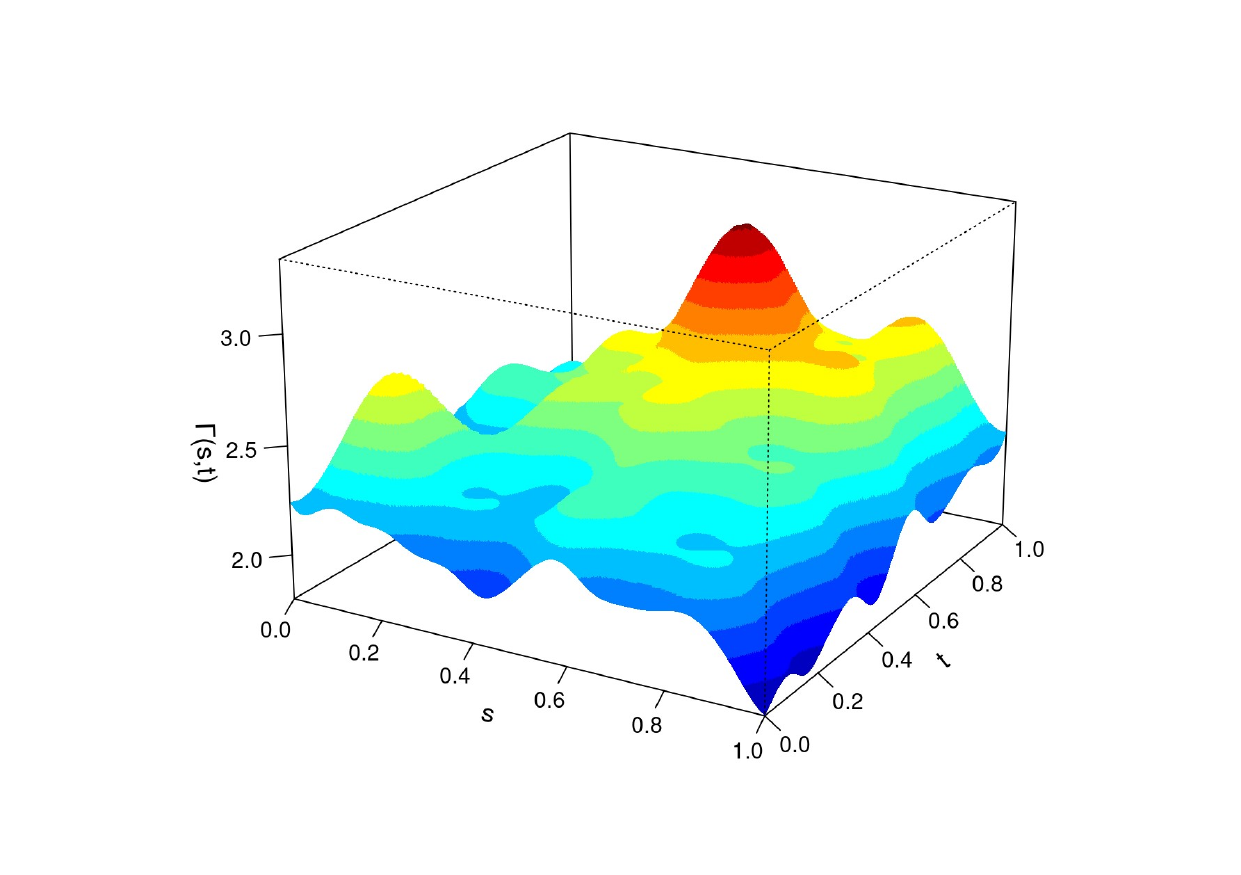}\hfill
\hspace{-0.8cm}\includegraphics[height = 0.25\textheight, width= .32\textwidth]{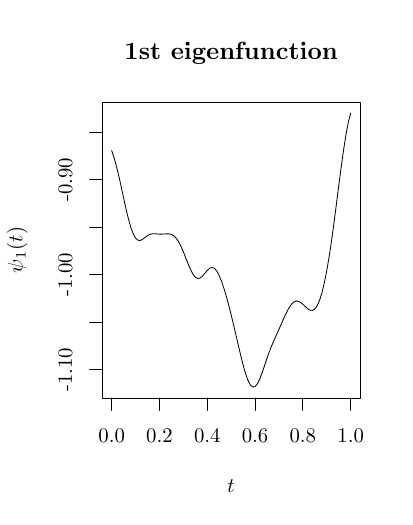}
\hspace{-0.6cm}\includegraphics[height = 0.25\textheight, width=.32\textwidth]{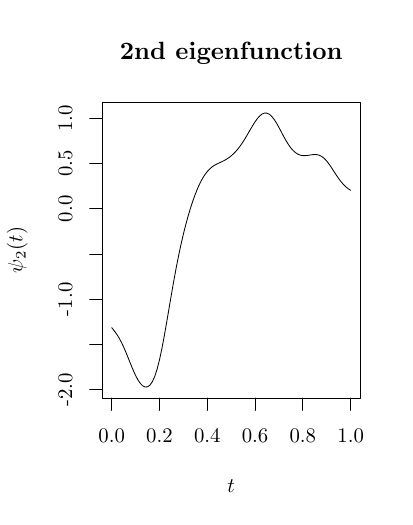}
\vspace{-.2cm}
\caption{\small  Simulation DGP: covariance function $\Gamma$ (left), the first two  eigenfunctions (middle and right) }
\label{fig:mean_cov_fun_plot}
\end{figure}

We consider 16 experimental setups, each consisting of 500 replications. They correspond to all combinations of the varying parameters  $N \in \{50, 100, 200\}$, $\mathfrak{m} \in \{25, 50, 100\}$, and $\sigma_0 \in \{0.25, 1\}$, except for the two combinations with $(N = 200, \mathfrak{m} = 100)$,  due to the excessive computational time required for the competing methods. The number of points $M_i$ are generated from a Poisson distribution with mean parameter $\mathfrak{m}$.  In all experiments, the sampling points $T_m^{(i)}$ were generated from a uniform distribution in $[0, 1]$, and the true eigen-elements used for computing the comparison measures were obtained by taking an accurate numerical eigen-decomposition of the true covariance function constructed. We focus on the first 9 eigenvalues, and the corresponding eigenfunctions, because they  account for more that  97.5\% of explained variance of the simulation DGP. See Table \ref{evalues-table} for  the eigenvalues and their cumulative explained variance. A plot of the covariance function along with its first 2 eigenfunctions can be seen in Figure \ref{fig:mean_cov_fun_plot}.

\subsection{Parameter settings, estimation and comparison measures}\label{sec_real_d}

With a simulated sample of pairs $(Y_m^{(i)}, T_m^{(i)})$, we proceed as indicated in Section \ref{sec:algorithm}, running twice the algorithm for the sake of self-consistency. 
To get the $\widetilde X^{(i)}$'s, we perform local constant kernel presmoothing of our curves at the points  $\{0, T_1^{(i)}, \dots, T_{M_i}^{(i)}, 1 \}$, and we linearly interpolate between these points. The bandwidth used for presmoothing is learnt by LS-CV on a randomly selected subset of 20 curves, where we selected the median of the 20 bandwidths. 
For the estimation of $\widehat H_t$ and $\widehat L_t$, as stated in Theorem \ref{thm:estimation-alpha_L} in the Appendix, a parameter $\gamma\in (0,1)$ determining the spacings $\Delta_*$ has also to be fixed.
This parameter,  which  defines the ``local neighborhood'' for estimating $H_t$ and $L_t$, is set equal to $0.75$. Slightly different other values we considered yielded very similar results. Then,  $\widehat H_t$ and $\widehat L_t$ is first computed on a grid $ \mathcal{T}_{param}\subset [0,1]$. To ensure enough points in the local neighborhoods when estimating the local regularity parameters, we adapt the size of the parameter estimation grid  such that $\#\mathcal{T}_{param} = [\mathfrak{m} / 3 ]$. Here, $[a]$ denotes the integer part of $a$. In the applications, $\mathfrak m$ is simply estimated by $\overline M$, the average of the $M_i$'s. We next complete the estimates $\widehat H_t$ and $\widehat L_t$ on a more refined grid $\mathcal{T}_{smooth}$ with 101 points,  by applying a smoothing splines procedure on the unit interval, with a data-driven number of knots  $[ \mathfrak{m} / 4 ]+1$. 
In Steps \ref{step3} and \ref{step3b} of the Algorithm, we start by computing $\widehat \nu$, $\widehat \sigma^2$ and $\widehat c_2$ on the grids $\mathcal{T}_{smooth}$ and $\mathcal{T}_{smooth}\times \mathcal{T}_{smooth}$, respectively. Finally,  in Step \ref{sze:tep5} where the diagonal correction is performed, we set   $\zeta = 0.1$.

Subsequently, the optimal bandwidth values $\widehat h^*(\lambda_j)$ and $\widehat h^*(\psi_j)$ were computed for each $1 \leq j \leq J= 9$, over a geometric grid $\mathcal H_N$  of 61 points, with 
$\min \mathcal H_N = \log(N) / (\overline M \sqrt{N})$  and  
$\max \mathcal H_N = 0.1 .$
The definition of the left end-point of $\mathcal H_N$ is in line with condition \eqref{eqdef:HN} on the bandwidth range. An upward correction of the bandwidths discussed at the end of Section \ref{sec:algorithm} is then applied in all our experiments to obtain the bandwidths $\widehat  h _{WPK}^*(\lambda _{j}) $
and $ \widehat  h _{WPK}^*(\psi _{j})$, which are used to compute the adaptive eigenvalues and eigenfunctions estimates, denoted $\widehat \lambda_{j,WPK}$ and $\widehat \psi_{j, WPK}$.

The bi-dimensional grid for computing the final covariance estimates is $\mathcal{T}_{smooth} \times \mathcal{T}_{smooth}$. Since the absolute value of the scalar product of any pair of our estimated eigenfunctions was less than $10^{-2}$, we do not use an orthogonality correction.

Our  estimates $\widehat \lambda_{j,WPK}$ and $\widehat \psi_{j, WPK}$ are then compared to the eigenvalue and eigenfunction estimates of \cite{Zhang2016}, obtained from the \textbf{R} package \texttt{fdapace}. See \cite{fdapaceR}. The estimation of the eigen-elements of \cite{Zhang2016} requires a bandwidth selection. Due to the exorbitant computational time of cross-validation (CV), we performed comparisons against the default bandwidth setting $b_{ZW, 2} = 0.1$ in the implementation \cite{fdapaceR}, corresponding to 10\% of the width of the sampling interval $\mathcal{T}$. We denote the corresponding eigen-element estimates as $(\widehat \lambda_{j, ZW, 2}, \widehat \psi_{j, ZW, 2})$.  Additional comparisons against a different default bandwidth $b_{ZW, 1} = 0.05$ can be found in the Supplementary Material.
As a measure of comparison for the eigenvalues, we use the ratio of absolute error
\begin{equation}\label{def:R_lam}
    \qquad \mathcal{R}_{ZW, 2}(\lambda_j) = \left|\widehat \lambda_{j,WPK} - \lambda_j\right|\Big/ \left|\widehat \lambda_{j, ZW, 2} - \lambda_j\right|.
\end{equation}
For the eigenfunctions, we use  the ratio of the $L^2-$norm errors, given by the square root of 
\begin{equation}\label{def:R_psi}
    \mathcal{R}_{ZW, 2}^2(\psi_j) = \frac{\int \left(\widehat \psi_{j, WPK}^c(t) - \psi_j(t) \right)^2 dt}{\int \left(\widehat \psi_{j, ZW, 2}^c(t) - \psi_j(t) \right)^2 dt}, \quad \text{where} \quad     \widehat \psi_j^c = \text{sign}\left(\langle \widehat \psi_j, \psi_j\rangle \right) \times \widehat \psi_j.
\end{equation}

\subsection{Empirical results} 
\label{sec:sim_results}

\begin{figure}[!ht]
\centering
\includegraphics[width=.24\textwidth]{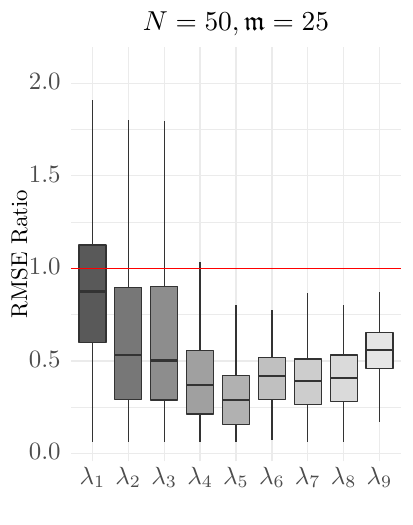}
\includegraphics[width= .24\textwidth]{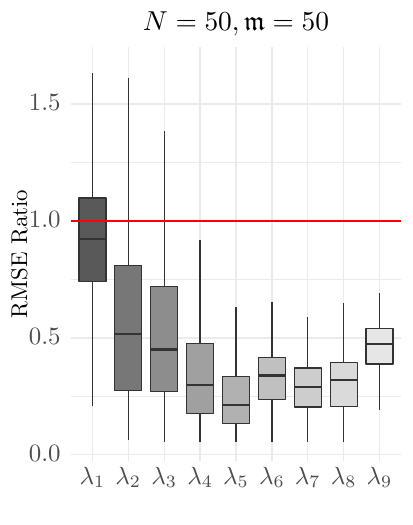}
\includegraphics[width= .24\textwidth]{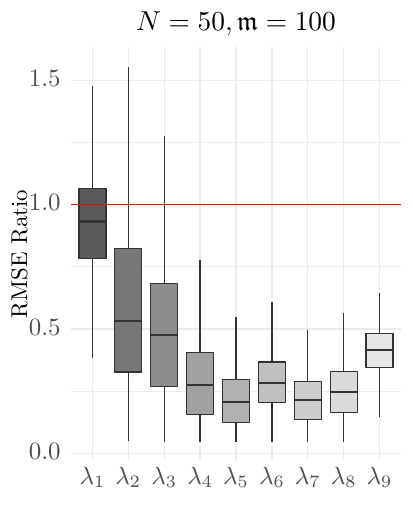}
\includegraphics[width= .24\textwidth]{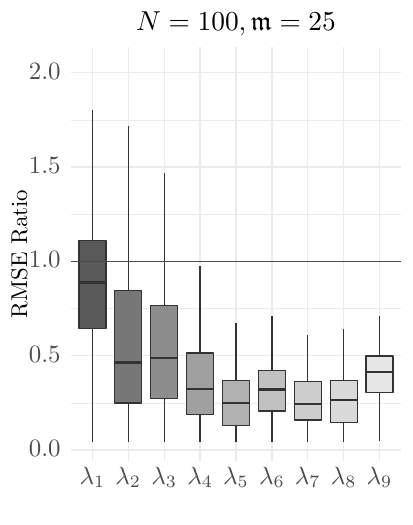} \hfill

\includegraphics[width= .24\textwidth]{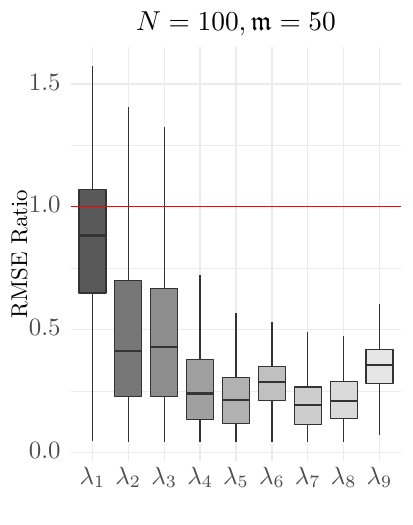}
\includegraphics[width= .24\textwidth]{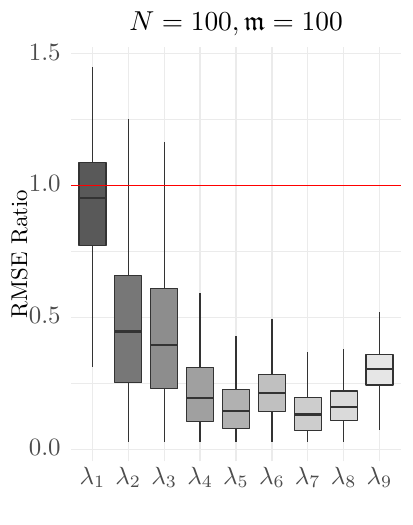}
\includegraphics[width= .24\textwidth]{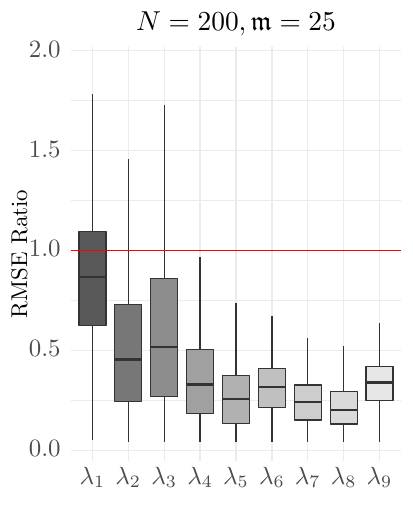}
\includegraphics[width= .24\textwidth]{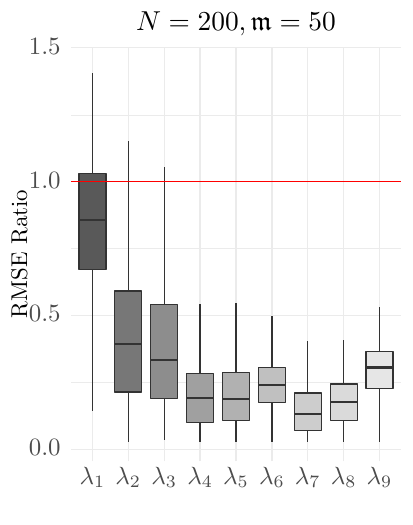}

\caption{\small Our method compared to the one in \cite{fdapaceR}:  the ratio $\mathcal{R}_{ZW, 2}(\lambda_j)$ of the absolute errors of  the eigenvalues estimates  for $\sigma_0 = 0.25$, $b_{ZW, 2} = 0.1$, with different values   $N$ (number of curves) and $\mathfrak{m}$ (average number  of random design points along each curve). Results from 500 replications.}
\label{fig:sigma_0.25_mfbm_0.1_val}
\end{figure}

\begin{figure}[!ht]
\centering
\includegraphics[width=.24\textwidth,height=4cm]{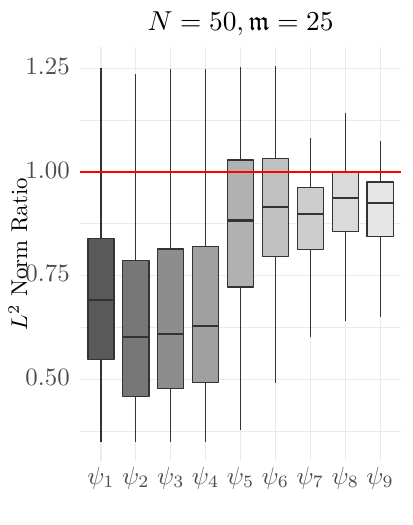}
\includegraphics[width=.24\textwidth,height=4cm]{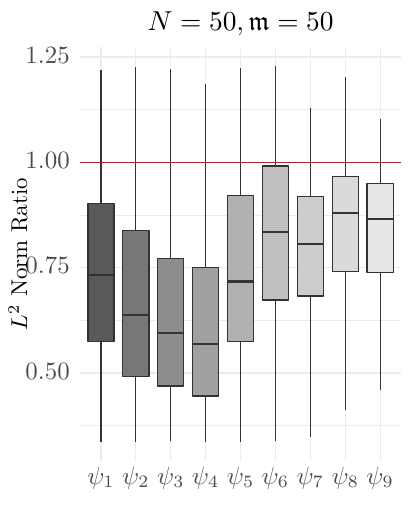}
\includegraphics[width=.24\textwidth,height=4cm]{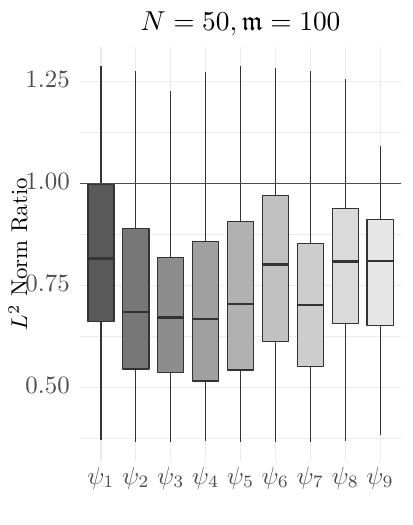}
\includegraphics[width=.24\textwidth,height=4cm]{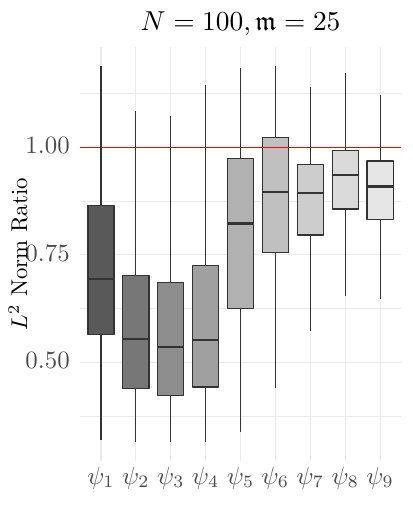} \hfill

\includegraphics[width=.24\textwidth,height=4cm]{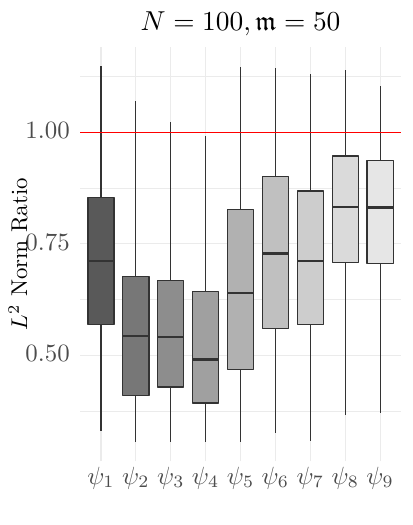}
\includegraphics[width=.24\textwidth,height=4cm]{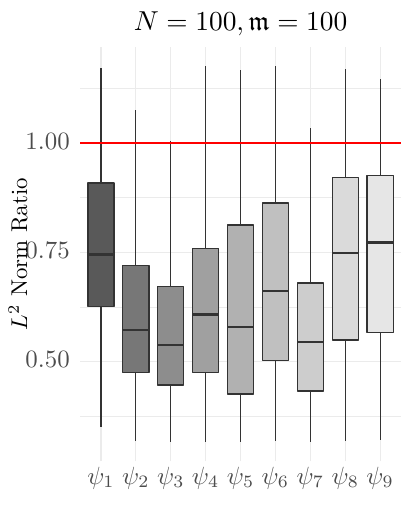}
\includegraphics[width=.24\textwidth,height=4cm]{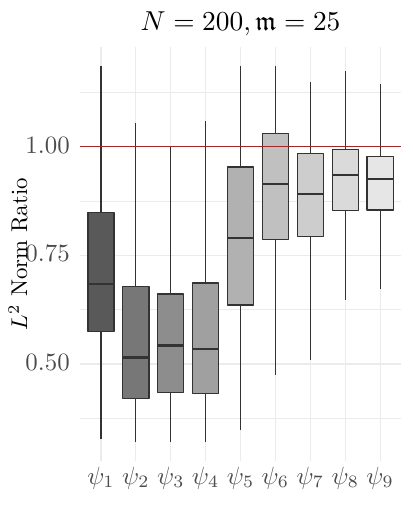}
\includegraphics[width=.24\textwidth,height=4cm]{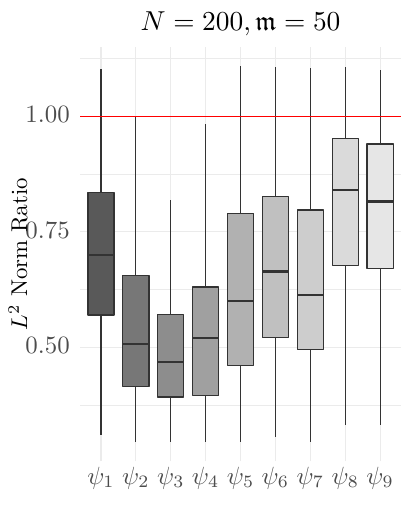}
\vspace{-.2cm}
\caption{\small  Our method compared to the one in  \cite{fdapaceR}:  the ratio $\mathcal{R}_{ZW, 2}(\psi_j)$ of the $L^2-$norm errors of the eigenfunctions estimates. The same simulation setup as in Figure \ref{fig:sigma_0.25_mfbm_0.1_val}.}
\label{fig:sigma_0.25_mfbm_0.1_fun}
\end{figure}

\begin{figure}[!ht]
	\vspace{-.4cm}
\centering
\includegraphics[width=.24\textwidth,height=4cm]{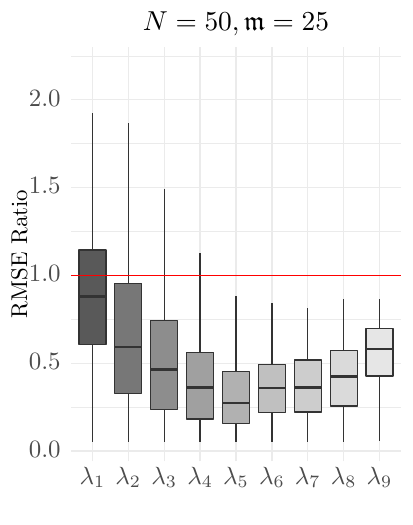}
\includegraphics[width=.24\textwidth,height=4cm]{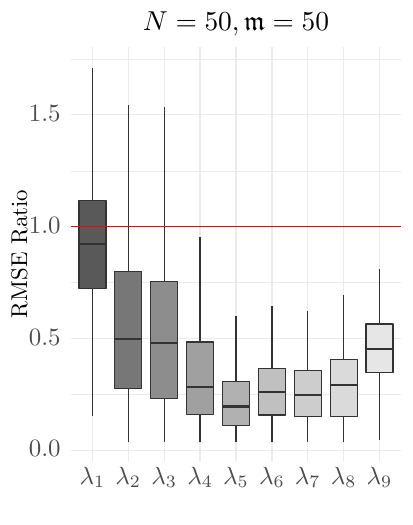}
\includegraphics[width=.24\textwidth,height=4cm]{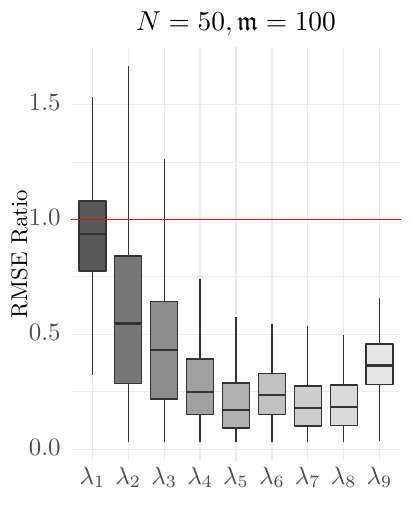}
\includegraphics[width=.24\textwidth,height=4cm]{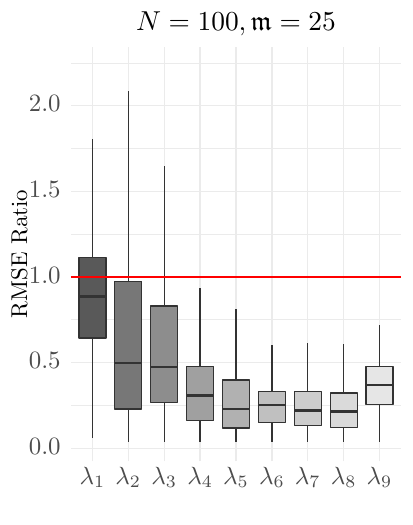} \hfill

\includegraphics[width=.24\textwidth,height=4cm]{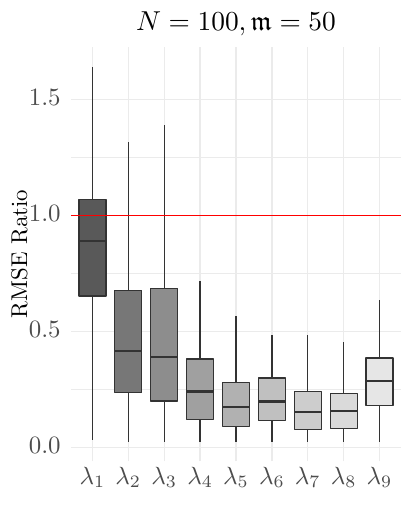}
\includegraphics[width=.24\textwidth,height=4cm]{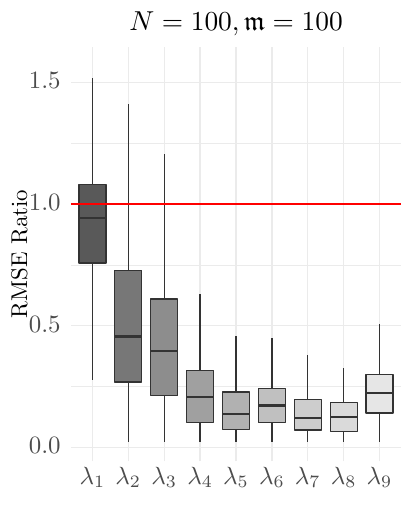}
\includegraphics[width=.24\textwidth,height=4cm]{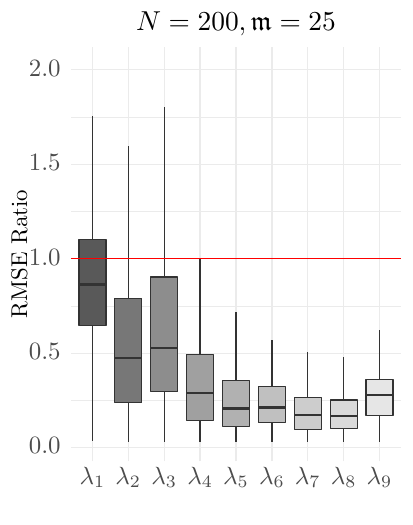}
\includegraphics[width=.24\textwidth,height=4cm]{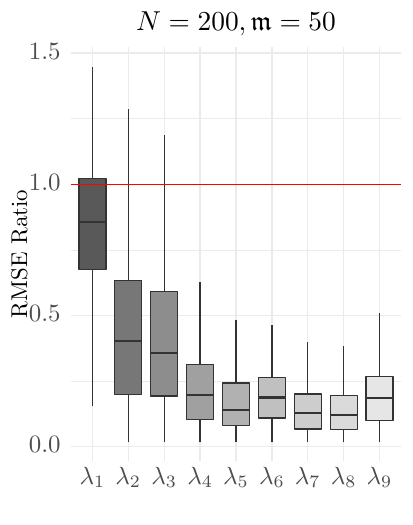}
\vspace{-.2cm}
\caption{\small Our method compared to the one in  \cite{fdapaceR}: the ratio $\mathcal{R}_{ZW, 2}(\lambda_j)$ of the absolute errors of  the eigenvalues estimates. The simulation setup as in Figure \ref{fig:sigma_0.25_mfbm_0.1_val}, with $\sigma_0 = 1$.}
\label{fig:sigma_1_mfbm_0.1_val}
\end{figure}

\begin{figure}[!ht]
\centering
\includegraphics[width=.24\textwidth,height=4cm]{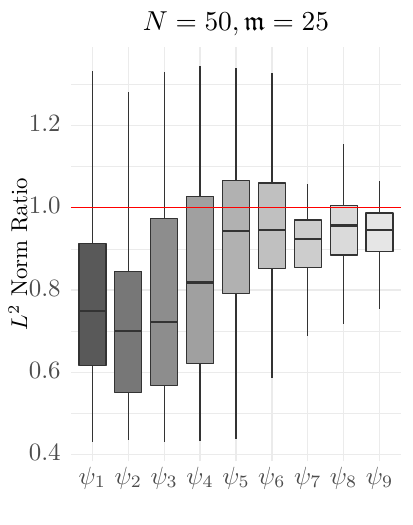}
\includegraphics[width=.24\textwidth,height=4cm]{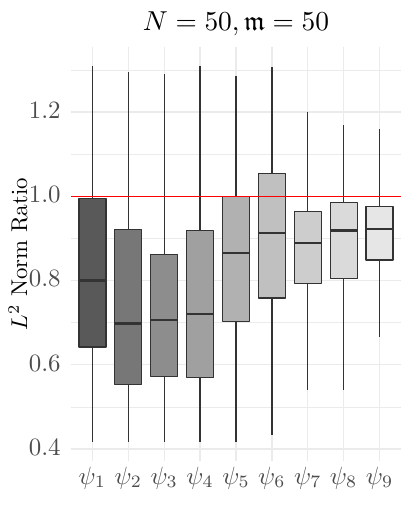}
\includegraphics[width=.24\textwidth,height=4cm]{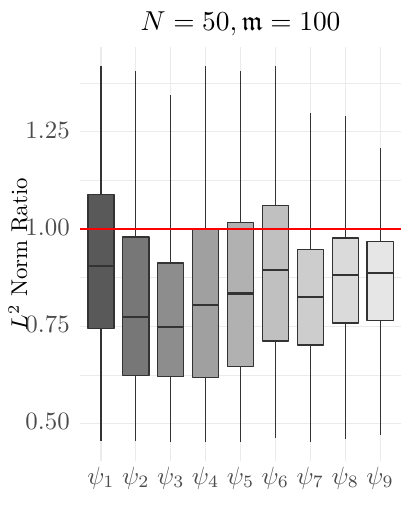}
\includegraphics[width=.24\textwidth,height=4cm]{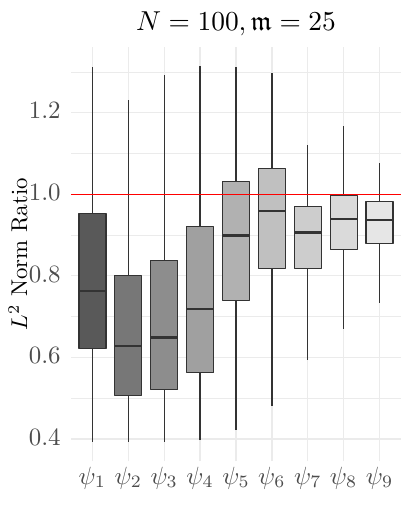} \hfill

\includegraphics[width=.24\textwidth,height=4cm]{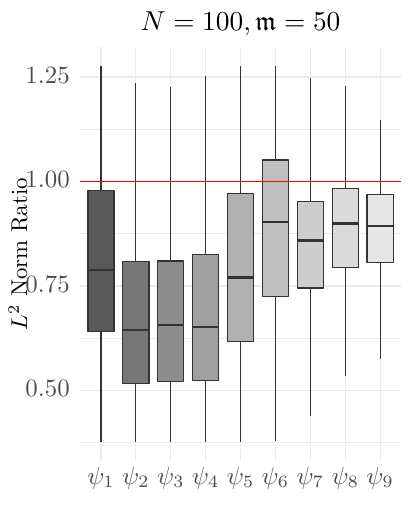}
\includegraphics[width=.24\textwidth,height=4cm]{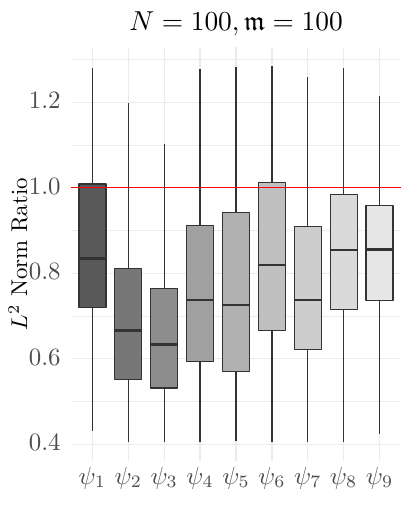}
\includegraphics[width=.24\textwidth,height=4cm]{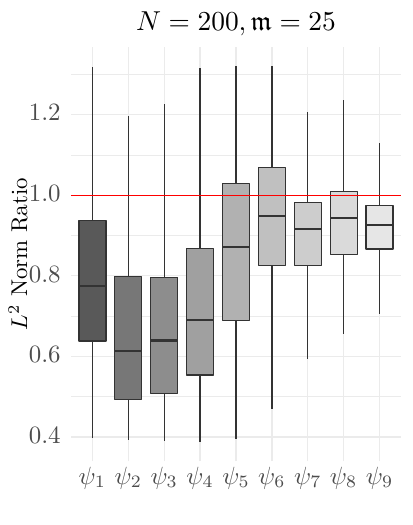}
\includegraphics[width=.24\textwidth,height=4cm]{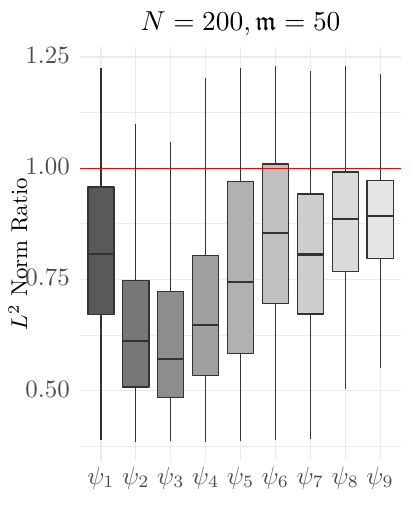}
\vspace{-.2cm}
\caption{\small Our method compared to the one in  \cite{fdapaceR}:  the ratio $\mathcal{R}_{ZW, 2}(\psi_j)$ of the $L^2-$norm errors of the eigenfunctions estimates. The simulation setup as in Figure \ref{fig:sigma_0.25_mfbm_0.1_fun},  with $\sigma_0 = 1$.}
\label{fig:sigma_1_mfbm_0.1_fun}
\end{figure}

The results on $\mathcal{R}_{ZW, 2}(\lambda_j)$ and $\mathcal{R}_{ZW, 2}(\psi_j)$ are presented in Figures \ref{fig:sigma_0.25_mfbm_0.1_val} to \ref{fig:sigma_1_mfbm_0.1_fun}. 
For clearer visualization of the plots, in all our boxplots, we remove values below the 5th quantile and above the 95th quantile, since the number of extreme values on each ends were roughly the same. It can be seen that our estimates of the eigen-elements perform favorably relative to \cite{fdapaceR}, where for virtually all setups, our error ratios are below 1, the benchmark for equal performance. Our estimates obtain up to 4 times less error for the eigenvalues, and almost 2 times less error for the eigenfunctions, demonstrating a big improvement. More empirical results are available in the Supplementary Material, which include comparisons to other  settings of \cite{fdapaceR}  (including a case where their bandwidth is selected by CV), lower signal-to-noise ratios, and a simpler simulation setup involving different constant Hurst functions for the fractional Brownian motion. We also present in the supplement  comparisons with the `first smooth, then estimate' approach using adaptive B-splines, and a setup with common design.   The conclusion remains consistent: our method outperforms competitors in most settings.

\begin{figure}[!htp]
	\centering
	\hspace{-.4cm}\includegraphics[width=0.5\textwidth, height=4cm]{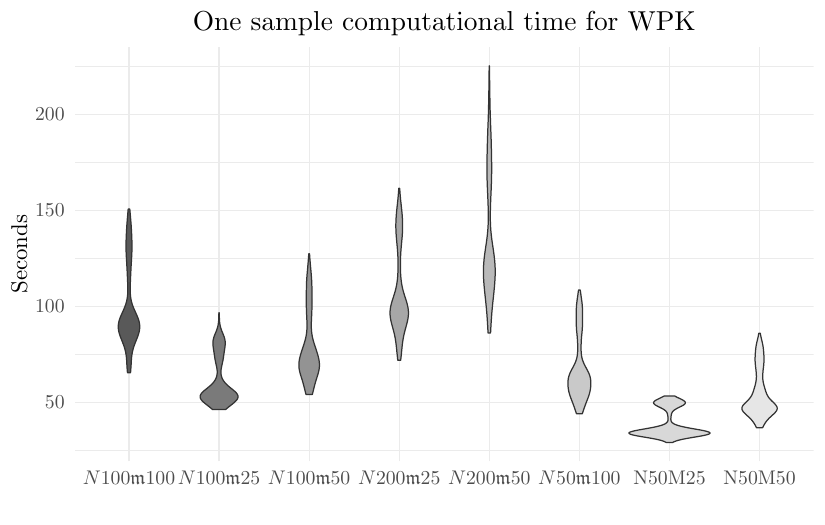}\hspace{-.1cm}
	\includegraphics[width=0.5\textwidth, height=4cm]{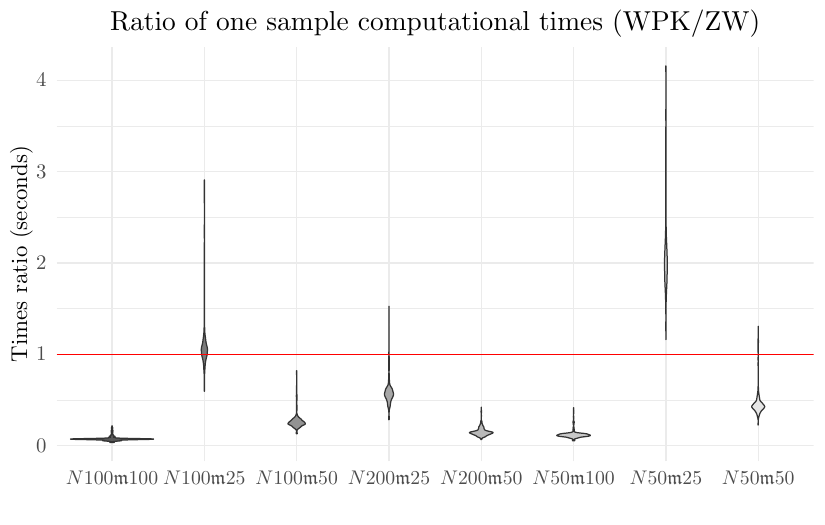}
	\vspace{-.2cm}
	\caption{\small  Our method (WKP) compared to the one in  \cite{fdapaceR} with fixed bandwidth (ZW):  violin plots of the computational times  \& the ratios for the simulated samples, for different $N$ and $\mathfrak m$ values.} 
	\label{fig:comp_sigma0.25}
	\vspace{-.2cm}
\end{figure}

Due to the explicit nature of our risk bounds, our bandwidth rule is also computationally efficient, even accounting for the various quantities that need to be estimated in calculations. As seen in Figure~\ref{fig:comp_sigma0.25}, our computational times are comparable for small sample sizes, in the sense that it does not take more than 100 seconds to run. However, for larger sample sizes, say with $N=100$ curves and $\mathfrak m = 100$ design points on average along each curve, our approach performs significantly faster by around an order of magnitude. Furthermore, this comparison is with respect to the default setting, plug-in bandwidths in \cite{fdapaceR}. With cross-validation instead of the default setting, the comparison of the computational times  will be significantly more favorable for us. 

\section{Real data application}\label{lets_see}
We return to the household electricity consumption data set, first seen in Section \ref{subsec:dgp_1}. In order to preserve the independence between curves and avoid issues with seasonality, we extracted the voltage curves in the month of June for analysis, corresponding to a summer month in France. (The analysis for January, a winter month, is provided in the Supplementary Material.) After removing curves with missing data points, the June sample resulted in $N=109$ curves with 1440 common design points. In order to perform comparisons with an independent design setup, we first draw the number of points along each curve $M_i$, $1\leq i \leq 109$, from a Poisson distribution with mean parameter $\mathfrak m \in \{25, 50\}$. Given the very irregular voltage curves, the values $N$ and $\mathfrak m$  correspond to a sparse setup, in the sense that $\mathfrak{m}^{2\underline H} \ll N$. Next, $M_i$ design points were then randomly sampled from each curve $i$ from a uniform distribution. Since the implementation of the pooling approach of \cite{Zhang2016}, as available in the package \cite{fdapaceR}, requires distinct $T_m^{(i)}$, $1\leq m \leq M_i$, $1\leq i \leq N$,  we sampled points $T_m^{(i)}$ from a continuous uniform distribution on $[0, 1]$ instead. For each curve $i$ and each $T_m^{(i)}$, the closest point to $T_m^{(i)}$ on the common design is then selected and its associated observed voltage value is recorded  as $Y_m^{(i)}$. 

\begin{figure}[ht]
\hspace{-.3cm}\includegraphics[scale=0.785]{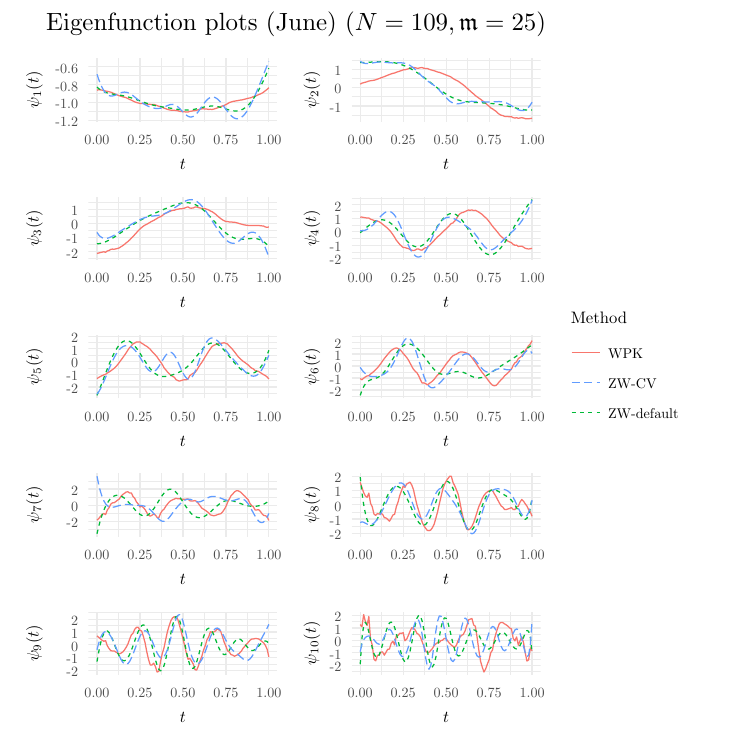}
\hspace{-1cm}\includegraphics[scale=0.785]{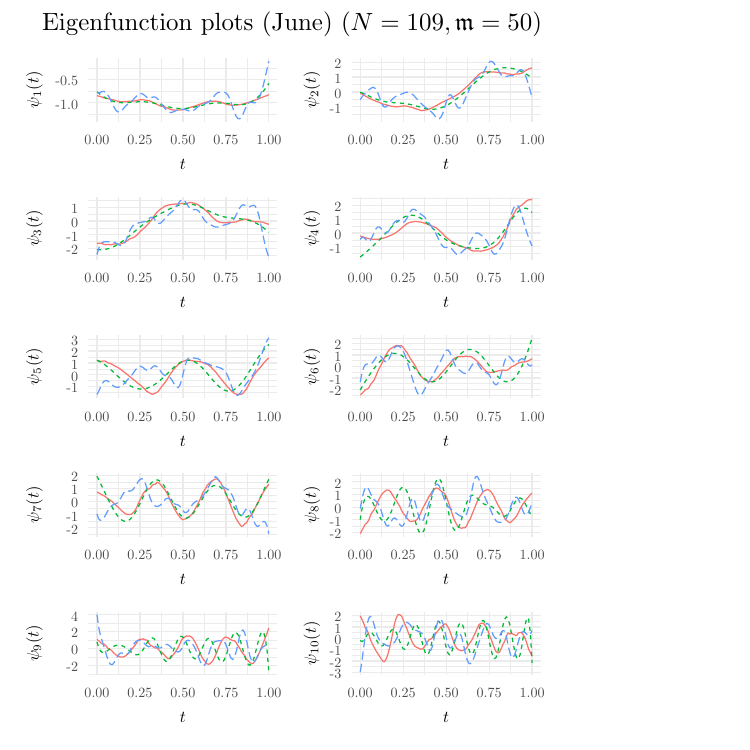}
\caption{\small Eigenfunctions estimates in the real data application: our method (WPK), the procedure from \cite{fdapaceR} with their cross-validation bandwidth (ZW-CV) and the default bandwidth (ZW-default). }
\label{efun_application_june}
\end{figure}

For our approach, we used  the same parameter settings as in the simulations, performing FPCA on a grid of 101 equally spaced points, with the same bandwidth and parameter grids. Comparisons were then made to the approach of \cite{Zhang2016} with both the  cross-validation bandwidth and  default bandwidth, and the results can be seen in Figure \ref{efun_application_june} and Table \ref{eval_june_m25}. The shapes of the eigenfunction estimates are somehow similar, but our approach yields smoother estimates, especially compared to the ZW-CV approach. The decrease of the eigenvalues estimates is quite similar too.

\begin{table}[ht]
\small \centering
\caption{\small  Eigenvalues estimates in the real data application: our method (WPK), the procedure from \cite{fdapaceR} with the generalized cross-validation bandwidth, (ZW-CV) and the default bandwidth (ZW-default). }
\label{eval_june_m25}
\begin{tabular}{l|c|cccccccccc}
Method & $\mathfrak m$ & $\lambda_1$ & $\lambda_2$ & $\lambda_3$ & $\lambda_4$ & $\lambda_5$ & $\lambda_6$ & $\lambda_7$ & $\lambda_8$ & $\lambda_9$ & $\lambda_{10}$ \\ 
  \hline
WPK & 25 & 3.27 & 0.31 & 0.18 & 0.06 & 0.04 & 0.01 & 0.00 & 0.00 & 0.00 & 0.00 \\ 
 & 50  & 3.35 & 0.28 & 0.21 & 0.11 & 0.11 & 0.06 & 0.02 & 0.02 & 0.00 & 0.01 \\ \hline 
  ZW-CV  & 25 & 3.66 & 0.33 & 0.29 & 0.10 & 0.06 & 0.03 & 0.02 & 0.01 & 0.01 & 0.00 \\ 
     & 50 & 3.35 & 0.35 & 0.25 & 0.21 & 0.16 & 0.12 & 0.09 & 0.06 & 0.05 & 0.03 \\ \hline
  ZW-default  & 25 & 3.63 & 0.31 & 0.20 & 0.03 & 0.02 & 0.01 & 0.00 & 0.00 & 0.00 & 0.00 \\ 
    & 50 & 3.20 & 0.21 & 0.15 & 0.08 & 0.03 & 0.02 & 0.01 & 0.00 & 0.00 & 0.00 \\ 
   \hline
\end{tabular}
\end{table}


\appendix
\section{Appendix}\label{sec:appendix_x}

Our approach adapts to the regularity of the process, a notion introduced in \eqref{eq:def_lr1x} and formally defined  in Section \ref{sec:def_loc_reg_x}. In Section \ref{ssec:est_x} we provide a new uniform concentration result for the regularity estimators.


\subsection{Local regularity in quadratic mean}\label{sec:def_loc_reg_x}

In the following,  $a\vee b$ and $a \wedge b$ denote the maximum and the minimum of $a$ and $b$, respectively. 
Let $H$ and $L$ be two functions defined on a compact interval $\mathcal T$ such that 
\begin{equation}\label{HL_bounds}
	0<\underline H := \min_{u\in\mathcal T} H_u  \leq  \max_{u\in\mathcal T} H_u  =: \overline H <1 \quad \text{and} \quad 0<\underline L := \min_{u\in\mathcal T} L_u  \leq  \max_{u\in\mathcal T} L_u  =: \overline L < \infty.
\end{equation}


\begin{definition}\label{def_global_reg}
	For given functions $H$ and $L$ 
	satisfying \eqref{HL_bounds}, the class $\mathcal{X}(H, L ;\mathcal T)$, or simply $\mathcal{X}(H, L)$, is the set of stochastic processes $X$
	satisfying the following conditions~: a constant $ \Delta_0 >0$ exists such that~: 
	
	\begin{enumerate}[({H}1)]
		\item\label{H:moments}    

		constants 
		$\mathfrak{a}>0$ and $\mathfrak{A}>0$ exist such that,
		for any  $p\geq 2$,
		\begin{equation*}
			\sup_{u,v\in\mathcal T, |u-v|\leq \Delta_0}  \EE\left[ | X_{u}-X_{v}|^{2p} \right] \leq
			\frac{p!}{2} \mathfrak{a} \mathfrak{A}^{p-2};
		\end{equation*}    
		
		\item\label{H:equivalent} constants $S\geq 0$ and  $\beta>0$ exist 
		such that, for any $u,v,t\in\mathcal T$ with $u\leq t \leq v$ and $|u-v|\leq \Delta_0$, 
		\begin{equation*}
			\left|
			\EE\left[ ( X_{u} -  X_{v})^{2} \right]
			-  L_t^2|u-v|^{2H_t}
			\right|
			\leq
			S|u-v|^{2H_t+2\beta}.
		\end{equation*}
		
	\end{enumerate}
	The function $ H$ defines  the local regularity of the process, while $L$ determines  the local Hölder constant.  
\end{definition}

Definition \ref{def_global_reg} extends the  definition of the local regularity proposed by \cite{Golovkine2021} to the whole domain $\mathcal T$. 
The condition \assrefH{H:moments}, which imposes sub-Gaussian increments, serves to derive the exponential bound for the concentration of the local regularity estimator using Bernstein's inequality for the squared increments. 


The classes $\mathcal{X}(H, L )$ are general, examples are provided in \cite{Golovkine2021}. More examples can be obtained by time deformation of processes in some $\mathcal{X}(H, L )$, see  Section \ref{sec:sim_design} above. 
The following result, for which the proof is given in the Supplementary Material,  provides few additional examples. 

\begin{lemApp}\label{lemma_mfbm2}
	Let $X\in \mathcal{X}(H_X, L_X  )$ and $Y\in \mathcal{X}(H_Y, L_Y )$, with continuous functions $(H_X,L_X)$, 
	$(H_Y,L_Y)$.
	\begin{enumerate}[(1).]
		\item  Assume a constant $C>0$ exists such that $\EE\left[ | X_{u}-X_{v}|^{4} \right] \leq C \EE^2\left[ | X_{u}-X_{v}|^{2} \right] $, $\forall u,v\in\mathcal T$. Let $\phi:\mathbb R \rightarrow \mathbb R$ be a  Lipschitz  continuous map with bounded derivative $\phi^\prime$. 
		Let $L_{\phi,t} = L_{X,t} \mathbb E^{1/2} [\phi^\prime (X_t)^2]$, $t\in \mathcal T$. If  $X\in \mathcal{X}(H_X, L_X )$, then $\phi(X)\in \mathcal{X}(H_X, L_\phi  )$.
		
		\item Assume that  $H_X  \leq  H_Y + \mathfrak b$, for some $\mathfrak b >0$. Then $X+Y \in \mathcal{X}(H_X,L_X)$.
		
		\item Assume that  $X$ and $Y$ are, zero-mean,  independent and $H_X = H_Y$. Then $X+Y \in \mathcal{X}(H_X,L_X+L_Y)$.
		
		\item Assume that  $H_X  \leq  H_Y + \mathfrak b$, for some $\mathfrak b >0$, $X$ and $Y$ are independent, and $Y$ is bounded. Then $XY \in \mathcal{X}(H_X, L )$, where $L_t =  L_{X,t} \mathbb E  [Y_t^2]$.
	\end{enumerate}
\end{lemApp}

\subsection{Regularity estimation}\label{ssec:est_x}

Let $X\in \mathcal{X}(H,L)$. The quality of  $\widehat H_t$, $\widehat L_t$ defined in \eqref{eq:hat-h_x}, depends on the quality of the presmoothing  estimators $\widetilde{X}_u$ of $X_u$. To quantify their behavior, we consider the uniform $L^p-$risk 
\begin{equation*}
	R_{\mathfrak m} (p) = \EE\left( \left|\widetilde{X}- {X}\right|_{\infty} ^{p} \right),
	\quad p\geq 1,
\end{equation*}
where $\mathfrak m$ is the expected number of observations on a curve. 
Here, and in the following, $|\cdot|_\infty$ is the uniform norm for the continuous  functions defined on $\mathcal T$. 
The  uniform risk above is averaged over the values of $M$, and therefore we let it depend on 
$\mathfrak m$. \color{black} Any type of nonparametric estimator $\widetilde{X}$ (local polynomials, splines, \emph{etc}) can be used, as soon as,  its uniform $L^p-$risk  is  suitably bounded for $p=2$ and $p>2$, respectively. 

To state our non-asymptotic uniform concentration result for the estimators $\widehat H_t$ and $\widehat L_t$, we need to define more precisely $t_1, t_2$ and $t_3$ used in \eqref{eq:hat-h_x}, in terms of $t\in\mathcal T$ and some small $\Delta_*>0$, namely when $t$ is close to 0 or 1. 
Let us consider $\underline t=0 \vee (t-\Delta_*)$, $\overline t = (t + \Delta_*) \wedge 1$, and
define the pair $(t_1,t_3)$ as follows: $(t_1,t_3)=(\underline t, \overline t)$ if $t\leq 1/2$ or $(t_1,t_3)= (\overline t, \underline t)$ otherwise. Then,  $t\in [t_1 \wedge t_2, t_1 \vee t_2] \subset [t_1 \wedge t_3, t_1 \vee t_3]$. 
Finally, 
consider the following mild condition~: a  constant $c$ exists such that
\begin{equation}\label{log_ratio_b}
	0< c^{-1} \leq  \log(N)/ \log(\mathfrak m) \leq c. 
\end{equation}

\begin{theorem}\label{thm:estimation-alpha_L}  
	Assume that  $X$ belongs to $\mathcal{X}(H,L)$, and $H$   and $L$ are Lipschitz continuous.  Moreover,  \eqref{log_ratio_b} holds true,  and positive constants $\mathfrak{c}$ and $\mathfrak{C}$ exist such that
	\begin{equation}\label{LP:1} 
		R_{\mathfrak m}(2p)
		\leq 
		\frac{p!}{2} \mathfrak{c} \mathfrak{C}^{p-2}, \qquad \forall p, \mathfrak m \geq 1.
	\end{equation}
	Assume also, constants  $ \tau>0, B>0$ exists such that 
	\begin{equation}\label{LP:rate} 
		R_{\mathfrak m}(2)  \leq B\Mmu^{-\tau},\quad \forall \mathfrak m  \geq 1. 
	\end{equation}
	Consider 
		$\epsilon = \log^{-\varrho}(\Mmu)$ and  $\Delta_*   =  \exp(-\log^{\gamma}(\Mmu))  $,   
	for some $ \varrho>1$ and $ 0 <\gamma <1 $.
	Then, for any $\Mmu$ larger than some constant $\Mmu_0$ depending on $B$, $\tau$, $\gamma$, $\varrho$, $H$, $\beta$, $L$, and for  some positive constants $\mathfrak{f} $ and $\mathfrak{g} $ we have
	\begin{equation}\label{eq:concentration-H}
		\PP\left( \left|\widehat H - H\right| _\infty > \epsilon
		\right)
		\leq
		\exp\left( -\mathfrak{f} N\epsilon^2 \Delta_*^{4\overline H} \right),
		\qquad 
		\PP\left( \left|\widehat L^2 - L^2\right| _\infty  > \epsilon \right)
		\leq
		\exp\left( - \mathfrak{g}N \epsilon^2 \frac{\Delta_*^{4\overline H}}{\log^2(\Delta_*) }    \right).
	\end{equation}
\end{theorem}

\medskip

Theorem \ref{thm:estimation-alpha_L} is proved in the Supplementary Material. A  condition like \eqref{LP:1} is satisfied by common estimators given the realization of  $M$. See for instance Theorem 1 in \citet{GAIFFAS2007} for the case of local polynomials. Condition \eqref{LP:rate} is also a mild uniform convergence condition satisfied by usual nonparametric regression estimators, given the number of points on a curve. See for instance \cite{Tsybakov2009} and \cite{BELLONI2015}. In particular, the required conditions for the uniform risk of $\widetilde{X}$ can be obtained under general forms of heteroscedasticity and mild conditons on the  distribution of $T$.   In order to guarantee that conditions \eqref{LP:1} and \eqref{LP:rate} remain true when taking expectation with respect to $M$, it suffices to impose a mild condition like $C \mathfrak m^b\leq M$ for some constants $b\in(0,1]$ and $C > 0$, which can reasonably be used for a wide panel of practical situations.  
Concerning the  quantities $R_{\mathfrak m}(2)$, $\Delta_* $ and $ \epsilon$, they are required to be such that, for some suitable $a>0$, $R_{\mathfrak m}(2)/\Delta^a_* + \Delta^{1/a}_*/\epsilon $ becomes negligible as $\Mmu$ increases. With  our choices for $\epsilon$ and $\Delta_*$, this holds true for any $a>0$. On the other hand, for the purpose of adaptive kernel smoothing in the FPCA context, under the mild condition that $\log(N)/\log(\mathfrak m)$ is bounded, as required in \eqref{log_ratio_b}, the effect of estimating $H$  is negligible as soon as $\epsilon$ is negligible compared to $\log^{-1}(\Mmu)$. This explains our condition $\varrho >1$. 
Finally, the condition $\gamma <1$ combined with the lower bound in \eqref{log_ratio_b} make  the  bounds for the concentration of $\widehat H $ and $\widehat L $ to be exponentially small when $N$ increases.  In conclusion, the only practical choice we have to make is that of $\gamma$, which was set equal to 0.75 in simulations.



\subsection{Assumptions}\label{sec_ap:ass}

\begin{assumpA}\label{ass_data}
The data generating process satisfies the following conditions. 	
\begin{enumerate}
     	\item\label{proc1} 
		The curves $X^{(i)}$ are independent realizations of $X$ which belongs to $\mathcal{X}(H,L,\mathcal T)$, the set of stochastic processes introduced in Definition \ref{def_global_reg}, where $\mathcal T$ is a compact interval, and $H$   and $L$ are Lipschitz continuous, bounded from below and from above, like in  \eqref{HL_bounds}.

        \item\label{data4_ass} The observations are the pairs $(Y_m^{(i)}, T_m^{(i)}) \in \mathbb{R} \times \mathcal{T}$, $1\leq m \leq M_i$, $1\leq i \leq N$, obtained according to  \eqref{data}. In the independent design case, \color{black} the $M_i$ are independent with mean $\mathfrak m$, and a constant $C_{\mathfrak m} >1$ (independent of $\mathfrak{m}$) exists such that
        \begin{equation}\label{frop01}
        	C_{\mathfrak m}^{-1} \mathfrak m\leq  M_i  \leq C_{\mathfrak m}\mathfrak m,\qquad \forall 1\leq i \leq N, 
        \end{equation} 
    	 and with a common design,  $M_1=\cdots=M_N = \mathfrak m$ and for any $1\leq m \leq \mathfrak m$,  $T_m^{(1)}= \cdots=T_m^{(N)}$. In both cases,  $N$ and $\mathfrak m$ satisfy \eqref{log_ratio_b}. Moreover, \color{black} $X^{(i)}$'s,  $M_i$'s and $T_m^{(i)}$'s are mutually independent. 
    	The conditional variance function $t\mapsto \sigma(t) >0$ is Lipschitz continuous, and the $e_m^{(i)}$'s are iid centered, sub-Gaussian variables with unit variance, and independent of $X^{(i)}$'s,  $M_i$'s and $T_m^{(i)}$'s.
        
        \item\label{ass_data_3} In the independent design case, the independent  $T_m^{(i)}$'s admit a  Lipschitz continuous density $f_T$ with Lipschitz constant $C_f$, and positive constants $C_{f,L}, C_{f,U}>0$ exist such that
        \begin{equation}\label{l_u_bdnf_ass}
        	C_{f,L} \leq f_T (t) \leq C_{f,U},\qquad \forall t\in\mathcal T.
        \end{equation}
\end{enumerate}	
\end{assumpA}

The condition \eqref{frop01} is a convenient technical condition which shortens the technical arguments required in the following, and has little practical relevance. It can relaxed by imposing probability bounds on the deviation of the $M_i$'s from the mean. Without any technical difficulty, condition \eqref{l_u_bdnf_ass} can be relaxed to allow  the density of the design points $T_m^{(i)}$ to be different for different curves. The condition \eqref{log_ratio_b} means that for the asymptotic results given in Section \ref{sec:th_grnd} above, we consider a triangular array of design points.

\begin{assumpA}\label{ass_pr_sm}
\begin{enumerate}
		\item The presmoothed curves $\widetilde X^{(i)}$, $1\leq i \leq N$, satisfy conditions \eqref{LP:1} and \eqref{LP:rate}.
	
\end{enumerate}	
\end{assumpA}

\begin{assumpA}\label{ass_ad_smo}
	The adaptive smoothing requires the following conditions. 
	\begin{enumerate}
\item The kernel $K$ is  non-negative, bounded,  supported on  $[-1,1]$, and $\inf K >0$ on a sub-interval of the support. 

		\item  The bandwidth range $\mathcal H_N$ is a set of positive numbers and a constant $\nu >1$ exists such that 
		\begin{equation}\label{eqdef:HN}
			N\left\{\mathfrak m \inf \mathcal H_N\right\}^2/\log^\nu(N) \rightarrow \infty \quad \text{and} \quad 
			\sup \mathcal H_N \rightarrow 0.
		\end{equation}
	\end{enumerate}	
\end{assumpA}

\subsection{Technical results}

The results in this section are the technical building blocks for the proofs of the main results. Their proofs  are provided in the Supplementary Material. 

\begin{lemApp}\label{lemma_Ih} Let $H \in(0,1)$ be a Lipschitz continuous function on $\mathcal T$, and 
		$I(h) = \int_0^1 D(t) h^{2H_t } dt $,  
	where   $D$ is some function such that $0< \underline D \leq D(t)\leq \overline{D} <\infty$. 
Then a constant $\underline  C$ exists such that 
\begin{equation}\label{eq:indet}
 0<\underline C \frac{ h^{2\underline H}}{\log(1/h) }
\leq 	I(h) \leq \overline D h^{2\underline H} ,\quad \forall h\in \mathcal H_N.
\end{equation}
Moreover, if $\widehat H$ is an estimator of $H$ with uniform concentration like in \eqref{eq:concentration-H} with $\epsilon \log (\mathfrak m) \rightarrow 0$,  then 
\begin{equation}\label{equiv:I_h}
\sup_{h\in\mathcal H_N} 	\left|\int_0^1 D(t) h^{2\widehat H_t } dt- I(h) \right| = o_{\mathbb P}(1).
\end{equation}
\end{lemApp}

\medskip

The lower bound in \eqref{eq:indet} cannot be narrowed without further assumptions on $H$. 
Property \eqref{equiv:I_h} shows that the rates of convergence are not deteriorated as soon as the estimator of $H$ uniformly concentrates faster than $\log^{-1} (\mathfrak m)$, which is a very mild condition. 

Let us next describe the behavior of the errors' conditional variance estimator defined in \eqref{eq:sigma-hat_x}.

\begin{lemApp}\label{WN_order_app}
	Assumption \ref{ass_data}.\ref{data4_ass} and \ref{ass_data}.\ref{ass_data_3}, and Assumption \ref{ass_ad_smo} hold true. 
	Then~: 
		\begin{enumerate}
		\item\label{lem1_42_o_app}  
		Constants $\underline b_W, \overline b_W\in(0,1]$ then exist  such that 
		\begin{equation}\label{sdyu1}
			\underline b_W N \leq \frac{ \mathbb E_{M} \left[\mathcal W_N(t;h) \right]}{\min\{1,\mathfrak m h\}} \leq \overline b_W N,\qquad  \forall h\in\mathcal H_N, \forall t\in\mathcal T.
		\end{equation}
		
		\item\label{lem1_42_n_app} 
		Constants $\underline c_W,\overline c_W\in(0,1]$ then exist  such that 
		\begin{equation}\label{sdyu2}
			\underline c_W N \leq \frac{ \mathbb E_{M} \left[\mathcal W_N(s,t;h) \right]}{\min\{1,(\mathfrak m h)^2\}} \leq \overline c_W N,\qquad  \forall h\in\mathcal H_N, |s-t|> 2h.
		\end{equation}
				
		\item\label{lem1_42_nd_app} 
		Constants $\underline d_W,\overline d_W\in(0,1]$ then exist  such that 
		\begin{equation}
			\underline d_W N \leq \frac{ \mathbb E_{M} \left[\mathcal W_N(s,t;h) \right]}{\min\{1,\mathfrak m^2 h|t-s|/2\}+ \min\{1,\mathfrak m (h-|t-s|/2)\}} \leq \overline d_W N, \quad \forall h\in\mathcal H_N,\; \forall|s-t|\leq  2h.
		\end{equation}
	
	\end{enumerate}
\end{lemApp}

\smallskip

\begin{lemApp}\label{lem:WN_conc_app}
 Assume that the conditions of Lemma A.\ref{WN_order_app} hold true.   Then
	\begin{equation}
		\sup_{h\in\mathcal H_N }  \sup_{s,t\in\mathcal T }  \left\{ 
		\left|  \frac {\mathcal W_N(s,t;h)}{ \mathbb E_{M} \left[\mathcal W_N(s,t;h) \right] } -1  \right| + 	\left|  \frac { \mathbb E_{M} \left[\mathcal W_N(s,t;h) \right] } {\mathcal W_N(s,t;h)}-1  \right| \right\} = o_{\PP} ( 1 ).
	\end{equation}
	
\end{lemApp}

\smallskip

\begin{lemApp}\label{N_gamma_order_app}
	Assume that the conditions of Lemma A.\ref{WN_order_app} hold true. Then~:
		\begin{enumerate}
		\item\label{lem_g_42_n_a7} 
		Constants $\underline c_{\mathcal N},\overline c_{\mathcal N}\in(0,1]$ then exist  such that 
		\begin{equation}\label{sdyu_g_app}
			\frac{\{1+o_{\mathbb P}(1)\} 	\underline c_{\mathcal N}}{N \min\{\mathfrak m h,(\mathfrak m h)^2\}}		
			\leq \frac{1}{   \mathcal N_\Gamma (s|t;h) } 
			\leq \frac{\overline c_{\mathcal N} \{1+o_{\mathbb P}(1)\}}{\max\{1,\mathfrak m h\}\times \EE[ \mathcal{W}_N(s,t;h)]},
			\qquad  \forall  |s-t|> 2h;
		\end{equation}
		with the $o_{\mathbb P}(1)$ uniform with respect to 	$h\in\mathcal H_N$;
		
		\item\label{lem_g_42_nd_8} 
		Constants $\underline d_{\mathcal N},\overline d_{\mathcal N}\in(0,1]$ then exist  such that 
		\begin{equation}\label{lem_g_49_nd_9}
			\frac{	\underline d_{\mathcal N} \{1+o_{\mathbb P}(1)\}}{\underline r(s,t;h)} \leq \frac{1}{\mathcal N_\Gamma (s|t;h)} \leq \frac{	\overline d_{\mathcal N} \{1+o_{\mathbb P}(1)\}}{\max\{1,\mathfrak m h\} \times \EE[ \mathcal{W}_N(s,t;h)]}, \quad \forall|s-t|\leq  2h,
		\end{equation}
		with the 	$o_{\mathbb P}(1)$ uniform with respect to 	$h\in\mathcal H_N$, where 
		$$
		\underline 	r(s,t;h)=
		\min\{\mathfrak m h, (\mathfrak m h)^2,\mathfrak m^2 h |t-s|/2\} +   \mathfrak m   \{h - |t-s|/2\}.
		$$
	\end{enumerate}
\end{lemApp}

\smallskip

\begin{lemApp}\label{lem:cvp_sigma}
	Assumption \ref{ass_data} holds true and  $N\mathfrak m b^2/\log(N) \rightarrow \infty$. Then 
	\begin{equation}\label{reta_sigb}
	 	\sup_{t\in\mathcal T} \left| \widehat \sigma^2(t;b) - \sigma^2(t) \right| = O_{\mathbb P}\left( b+ b^{2\underline H} + \{\log(N)/[N\min(1,\mathfrak m b^2)]\}^{1/2} \right),
	\end{equation}
with defined $\widehat \sigma^2(t;b)$ in \eqref{eq:sigma-hat_x}.
In the common design case, for any $b >  C^{-1}_d \mathfrak m^{-1}$,   with $C_d$ in \eqref{sp_cd}, we have
	\begin{equation}\label{reta_sigb_c}
	\sup_{t\in\mathcal T} \left| \widehat \sigma^2(t;b) - \sigma^2(t) \right| = O_{\mathbb P}\left( b+b^{2\underline H} + \{\log(N)/N\}^{1/2} \right).
	\end{equation}
\end{lemApp}

\smallskip

The next result concerns the diagonal correction \eqref{diag_corr_eq}. Let 
\begin{equation}\label{diag_corr_eq_hat}
	\Hat{\Hat{d}}_N(s,t;h)=	
	\frac{\widehat \sigma(s;b) \widehat \sigma(t;b)}{\mathcal{W}_N(s,t;h)}\sum_{i=1}^N w_i(s;h)w_i(t;h) \sum_{m=1}^{M_i}W_m^{(i)}(s;h)W_m^{(i)}(t;h),
\end{equation}
with $\widehat \sigma^2(t;b)$ defined in \eqref{eq:sigma-hat_x}. Note that, by  construction, $\Hat{\Hat{d}}_N(s,t;h)=\widehat d_N(s,t;h)=0$ when $|s-t|>2h$.

\begin{lemApp}\label{repair_diag}
Assume the conditions of Lemma A.\ref{lem:cvp_sigma} hold true. Moreover,  $\sigma(s)\sigma(t)$ is estimated by $\widehat \sigma(s;b) \widehat \sigma(t;b)$ with $b = h^a$ for some $\max(1/2, 2\underline H -1) < a < 1$. Let $\widetilde \Gamma$ and $\widehat \Gamma$ be defined as in \eqref{eq_def_empG} and \eqref{cov_est_corr}, respectively.  Then 
\begin{equation}\label{ine_repari_d1}
\iint_{\mathcal T\times \mathcal T}\left\{\widehat d_N(s,t;h)- \Hat{\Hat{d}}_N(s,t;h) \right\}^2 dsdt \leq  o_{\PP}(1) \times \iint_{\mathcal T \times \mathcal T}\left\{\widehat \Gamma (s,t;h) - \widetilde \Gamma (s,t)\right\}^2 dsdt,
\end{equation}
and the $o_{\PP}(1)$ term does not depend on $h\in\mathcal H_N$. 
\end{lemApp}

\medskip

Theorems A.\ref{th:theorem1_app} and A.\ref{th:theorem2_app} below are the cornerstone for justifying Theorem \ref{thm:thm-1}. They provide Taylor expansion (representation) for each eigen-element. Theorems A.\ref{th:theorem1_app} and A.\ref{th:theorem2_app} are versions of  independent  interest of \cite[equations (2.8) and (2.9)]{Hall2006c},  \cite{HallNasab2009}; see also \cite{Jirak_Wahl}. 
They provide a representation between the eigen-elements of $\Gamma$ and those of the (infeasible) empirical estimator $\widetilde \Gamma$. Here, we provide an extended version of the results of \cite{HallNasab2009}, such that we can derive the Taylor expansions with more general estimators of the covariance function.


Consider the following notation: for  $M$ a square-integrable functions of two variables on $\mathcal T \times \mathcal T$, write 
$$
\N M \N^2 = \iint_{\mathcal T \times \mathcal T} M^2(s,t)dsdt= \iint M^2.
$$

For the covariance function $\Gamma$, assumed to be continuous on $\mathcal T \times \mathcal T$, we can write the spectral decomposition 
\begin{equation}\label{Eq:k-spectral}
	\Gamma (u,v) = \sum_{j=1}^\infty \lambda_j \psi_j(u) \psi_j(v), 
\end{equation}
where $\psi_1, \psi_2, \dots $ denotes a complete orthonormal sequence of continuous  eigenfunctions in $\mathcal{C}$, corresponding to the respective eigenvalues $\lambda_1\geq \lambda_2 \geq \ldots \geq 0$. Let 
\begin{equation} \label{def:eq_rho}
	\rho_1 =	\lambda_1 - \lambda_2 >0 \quad \text{and}\quad 	\rho_j = \min (\lambda_j - \lambda_{j+1}, \;\lambda_{j-1} - \lambda_j) >0, \quad \forall j\geq 2, 
\end{equation}
and
\begin{equation}\label{xi_eta_def_app}
	\xi_j = \inf \{1 - (\lambda_k/\lambda_j): \lambda_k < \lambda_j\} \;\;\text{ and } \; \;
	\eta_j = \inf \{(\lambda_k/\lambda_j)-1: \lambda_k >\lambda_j\}.
\end{equation}
Let $L$ denote a second symmetric kernel with associated non negative eigenvalues sequence $\theta_1 \geq \theta_2 \geq \dots$ and the complete orthonormal sequence of eigenfunctions $\phi_1, \phi_2,\dots$, such that we have
\begin{equation}\label{Eq:L-spectral}
	L(u,v) = \sum_{j=1}^\infty \theta_j \phi_j(u) \phi_j(v).
\end{equation}

\begin{thApp}\label{th:theorem1_app}
	Let $\Gamma$ and $L$ be two operators as in \eqref{Eq:k-spectral} and \eqref{Eq:L-spectral}. Then, for each $j\geq 1 $ for which
	\begin{equation} 
		\rho_j >0 \qquad \text{ and } \qquad 	\N L - \Gamma \N   \leq \frac{1}{2}\lambda_j \min(\xi_j, \eta_j),
	\end{equation}
	we have
	\begin{equation} 
		\theta_j - \lambda_j  =  \iint_{\mathcal T \times \mathcal T} (L-\Gamma) (s,t) \psi_j (s)\psi_j(t) dsdt + \mathfrak L_j, \qquad \text{with} \quad  \left| \mathfrak L_j \right|  \leq C_\lambda \times  \N L - \Gamma \N ^2,
	\end{equation}
	for some constant $C_\lambda>0$ depending on $\Gamma$, $\lambda_j$,  $\rho_j$, $\xi_j$ and $\eta_j$, but not on $L$.
\end{thApp}

\smallskip

\begin{thApp}\label{th:theorem2_app}
	Assume that the conditions of Theorem A.\ref{th:theorem1_app}  hold true. Then
	\begin{equation}\label{thm1-psihat-bound}
		\phi_j(t) - \psi_j(t) = \sum_{k:k\neq j}(\lambda_j - \lambda_k)^{-1}\psi_k(t) \iint (L-\Gamma) (u,v) \psi_j (u)\psi_k(v)dudv + \mathfrak{S}_j(t),\qquad t\in\mathcal T,
	\end{equation}
	and 
	$ \|\mathfrak{S}_j \|_2 \leq C_\psi \times  \N L - \Gamma \N ^2 	$
	for some constant $C_\psi>0$ depending on $\Gamma$, $\lambda_j$, $\rho_j$, $\xi_j$ and $\eta_j$, but not on $L$.
\end{thApp}

\subsection{Main results: outline of the proofs}

In the following, $C,c,...$ denotes constants with possibly different values at different occurrences. Recall that the symbol $\lesssim$ means that the left side is bounded by a constant times the right side. Meanwhile, the symbol $\asymp$ means the left side is bounded above and below by constants times the right side.

\begin{proof}[Proof of Theorem \ref{thm:thm-1}]
For this result, all the unknown quantities appearing in the risk bounds \eqref{eq:evalue-bound_x} and \eqref{eq:efunction-bound_x}, as well in the diagonal correction 
	$\widehat d_N(s,t;h)$, defined in \eqref{diag_corr_eq} and used to build $\widehat \Gamma(s,t;h)$, are supposed given. Let us consider another infeasible covariance function estimator, that is 
	\begin{equation}\label{cov-fct_W}
		\widetilde \Gamma_W(s,t;h) = \frac{1}{\mathcal{W}_N(s,t;h)}\sum_{i=1}^N w_i(s;h)w_i(t;h)\left\{ X_s^{(i)} -  \mu_{W}(s;h) \right\} \left\{ X_t^{(i)} - \mu_{W}(t;h)\right\},\quad s,t\in\mathcal T,
	\end{equation}
	with
		$\mu_{W} (t;h) =  \mathcal{W}_N(t;h)^{-1}\sum_{i=1}^N w_i(t;h) X_t^{(i)}$.
	It is the empirical covariance function computed with the curves as selected for building $\widehat \Gamma(s,t;h)$. Let $\widetilde \lambda_{j,W}$ and $\widetilde \psi_{j,W}$ denote the $j-$th eigenvalue and the corresponding eigenfunction of the covariance operator defined by $\widetilde \Gamma_W(s,t;h)$.
	Let $\widehat \lambda_{N,j}$ and $\widehat \psi_{N,j}$  denote the $j-$th eigenvalue and eigenfunction obtained with $\widehat \Gamma_N(s,t;h)$ defined in \eqref{cov-function_x}, the uncorrected for diagonal bias estimator of $\Gamma(s,t)$. 
	Weyl's inequality and the last part of Lemma A.\ref{repair_diag} guarantees that 
	$$
	\EEMT\left[\left\{\widehat \lambda_{N,j} - \widehat \lambda_{j}\right\}^2\right] \leq Ch^2,
	$$
	for some constant $C$ depending only on the  function $\sigma^2(\cdot)$ and the length of $\mathcal T$. 
	We then decompose 
	$$
	\widehat \lambda_{j}  - \widetilde \lambda_j = \{\widehat \lambda_{j} - \widetilde \lambda_{W,j} \}  + \{ \widetilde \lambda_{W,j} - \widetilde \lambda_j\} \quad \text{ and } \quad  \widehat \lambda_{j} - \widetilde \lambda_{W,j}  =  \{\widehat \lambda_{j}  - \widehat \lambda_{N,j} \} + \{\widehat \lambda_{N,j}  - \widetilde \lambda_{W,j} \} ,
	$$
and, using Theorem A.\ref{th:theorem1_app} and Lemma A.\ref{repair_diag}, we  derive the representations
\begin{equation}\label{Taylor_W_hat}
	\widehat \lambda_{j} - \widetilde \lambda_{W,j} =  \iint (\widehat \Gamma - \widetilde \Gamma_W )  \psi_j\psi_j + R_{\lambda,1}\qquad \text{ and } \qquad \widetilde \lambda_{W,j} - \widetilde \lambda_j  =   \iint (\widetilde \Gamma_W - \widetilde \Gamma ) \psi_j \psi_j + R_{\lambda,2}.
\end{equation}	
The remainder terms $R_{\lambda,1}$ and $R_{\lambda,2}$ are  negligible compared to the double integrals. Next, since $(a+b)^2 \leq 2a^2 + 2b^2$, we show that, up to negligible terms,
	$$
	 \EEMT\left[\left\{\widehat \lambda_j - \widetilde \lambda_{j,W}\right\}^2\right] \qquad \text{and} \qquad \EEMT\left[\left\{\widetilde \lambda_{j,W} - \widetilde \lambda_{j}\right\}^2\right] ,
	$$
are bounded by $\mathcal{B}_{1,N} (\widehat \lambda_j ;h) + {B}_{2,N} (\widehat \lambda_j ;h)$ (squared bias + variance terms) and $  \mathcal{B}_{3,N} (\widehat \lambda_j ;h)$ (penalty term), respectively. 
To derive the expression of $\mathcal{B}_{1,N} (\widehat \lambda_j ;h) + {B}_{2,N} (\widehat \lambda_j ;h)$, we write $\widehat X^{(i)} _t(h)= X_t^{(i)}+ B_t^{(i)}(h)+V_t^{(i)}(h)$, with
	$$ 
	B_t^{(i)}(h) =  \EEMT [\widehat X_t^{(i)}(h)] - X_t^{(i)} \quad \text{ and } \quad V_t^{(i)}(h) = \widehat X_t^{(i)}(h) - \EEMT [\widehat X_t^{(i)}(h)],\quad t\in\mathcal T,
	$$
which are the bias and the stochastic part of $\widehat X_t^{(i)}(h)$, respectively. 
	The bias term depends on the $X^{(i)}$, $M_i$ and the $T_m^{(i)}$'s, while the stochastic term depends on $M_i$, $e_m^{(i)}$ and $T_m^{(i)}$. For deriving the expression of $\mathcal{B}_{1,N} (\widehat \lambda_j ;h)$ we need to bound the expectation of the square of $B_t^{(i)}(h)$, which depends on the functions $H$ and $L$. The details are provided in Section \color{red}V \color{black} in the Supplementary Material.

For the eigenfunctions, using Theorem A.\ref{th:theorem2_app}, we derive the expansions
\begin{equation}\label{pory1}
	\widehat \psi_j(t) -\widetilde \psi_{W,j}(t) = 
	 \sum_{k:k\neq j}(\lambda_j - \lambda_k)^{-1}\psi_k(t) \iint (\widehat \Gamma - \widetilde \Gamma_W )  \psi_j \psi_k + R_{\psi,1}(t),
\end{equation}
and
\begin{equation}\label{pory2}
 \widetilde  \psi_{W,j}(t) -  \widetilde  \psi_{j}(t) = 
  \sum_{k:k\neq j}(\lambda_j - \lambda_k)^{-1}\psi_k(t) \iint (\widetilde \Gamma_W - \widetilde \Gamma)  \psi_j \psi_k + R_{\psi,2}(t).
\end{equation}
The $L^2-$norms of the remainders $R_{\psi,1}$ and $R_{\psi,2}$ are easily shown to be negligible compared to the $L^2-$norms of the sums of double integrals in \eqref{pory1} and \eqref{pory2}, respectively. We then follow the same lines like for the eigenvalues, and derive the three terms of the risk bound, that are squared bias, variance terms and penalty term, respectively. For practical purposes, the calculations are done for  versions of the representations  \eqref{pory1} and \eqref{pory2} with truncated sums.  Since the truncation error can be arbitrarily small, the rates of convergence of the estimates are not altered.  More precisely, for any $\mathcal K = \{1,\ldots,K_0\}$, we can write 
$$
\sum_{k:k\neq j} = \sum_{k\in\mathcal K :k\neq j} + \sum_{k\not \in \mathcal K:k\neq j} = \sum_{k\in\mathcal K :k\neq j} + \text{ truncation error}.
$$
Since the $\sum_{k:k\neq j}$ is convergent, for any $C>0$, an integer $K_0$ exists such that the norm of the truncation error  is smaller than $C$ times the norm of the complete sum $\sum_{k:k\neq j}$. The detailed derivations are provided in Section \color{red}IV \color{black} in the Supplementary Material. It also follows from the arguments provided there that 
replacing  $\psi_j$ and $\psi_k$ with the constant function equal to 1 when computing  $ {\mathcal{B}}_N(\widehat\lambda_j ;h) $ and $ {\mathcal{B}}_{N} (\widehat\psi_j ;h)$, may change the constants but does not change the rates of convergence of the bounds. This, combined with Corollary \ref{corr_rates_est} or \ref{corr_rates_est_cd}, guarantee that running the Algorithm in Section \ref{sec:algorithm} with the \emph{ad-hoc} proxy of $\psi_j$ equal to the constant function equal to 1, leads to estimates $\widehat \lambda_j$ and $\widehat \psi_j$ with the same rates of convergence.  \end{proof}


\begin{proof}[Proof of Corollary \ref{corr_rates_est}]
Lemmas  A.\ref{WN_order_app}, A.\ref{lem:WN_conc_app} and A.\ref{N_gamma_order_app} imply that the double integrals in the expressions of 
$\mathcal{B}_{2,N} (\widehat \lambda_j ;h)$, $\mathcal{B}_{3,N} (\widehat \lambda_j ;h)$, $\mathcal{B}_{2,N} (\widehat \lambda_j ;h)$ and $\mathcal{B}_{3,N} (\widehat \lambda_j ;h) $ can be restricted to the domain $\{|s-t|>2h\}$, because the integration  over the set $\{|s-t|\leq 2h\}$ is negligible. Moreover, a constant $C>0$ exists such that 
$$
\frac{C^{-1}\{1+o_{\mathbb P}(1)\}}{N\min\left\{\mathfrak{m}h, (\mathfrak{m}h)^2\right\}} \leq \mathcal{B}_{2,N} (\widehat \lambda_j ;h) + \mathcal{B}_{3,N} (\widehat \lambda_j ;h) \leq \frac{C \{1+o_{\mathbb P}(1)\}}{N\min\left\{\mathfrak{m}h, (\mathfrak{m}h)^2\right\}},
$$
and similar bounds can be derived for 
$\mathcal{B}_{2,N} (\widehat \psi_j ;h) + \mathcal{B}_{3,N} (\widehat \psi_j ;h)$. On the other hand, the squared bias terms $\mathcal{B}_{1,N} (\widehat \lambda_j ;h)$ and $\mathcal{B}_{1,N} (\widehat \psi_j ;h)$ can rewritten under the form
$\int B(t)h^{2H_t} dt $,
with some specific function $B$  determined by $L$, $K$, $\sigma^2$, $\nu$, $c_2$ and the true eigen-elements. Thus, to find the minima of the risk bounds for $\widehat \lambda_j $ and $\widehat \psi_j$ with respect to $h$, we need to find values $h$ such that  
$$
h^{2\underline H-1} \int B(t) h^{2H_t-2\underline H} \; dt \asymp \frac{1}{N\min\left\{\mathfrak{m}h^2, \mathfrak{m}^2 h^3 \right\}}.
$$
Let $h^*$ be such a solution, and note that by construction $\log(1/h^*)\lesssim \log(N\mathfrak{m})$.
 Lemmas A.\ref{lemma_Ih} then indicates 
	that  $h^*$ either  satisfies 
	\begin{equation}
		\left( N\mathfrak{m} \right)^{- \frac{1}{2\underline H + 1}} \lesssim h^* \lesssim \left( \log(N\mathfrak m)/ N\mathfrak{m} \right)^{\frac{1}{2\underline H + 1}} \qquad \text{ or } \qquad 
	\left( N\mathfrak{m}^2\right)^{- \frac{1}{2\underline H + 2}} \lesssim h^* \lesssim \left( \log(N\mathfrak{m})/N\mathfrak{m}^2\right)^{\frac{1}{2\underline H + 2}},
\end{equation}
depending on the range of $h$, that is either we have 
 $\min\left\{\mathfrak{m}h^2, \mathfrak{m}^2 h^3 \right\} = \mathfrak{m}^2h^3$ or 
  $\min\left\{\mathfrak{m}h^2, \mathfrak{m}^2 h^3 \right\} = \mathfrak{m}^2h^3$, respectively. 
	Gathering facts, we have that 
\begin{equation}
	\max\left[ \left( N\mathfrak{m}  \right)^{- \frac{1}{2\underline H + 1}},  \left( N\mathfrak{m}^2\right)^{- \frac{1}{2\underline H + 2}}\right] \lesssim h^* \lesssim \max\left[\left(\log(N\mathfrak{m})/N\mathfrak{m} \right)^{\frac{1}{2\underline H + 1}},  \left( \log(N\mathfrak{m})/N\mathfrak{m}^2\right)^{\frac{1}{2\underline H + 2}}\right].
\end{equation}
Substituting $h^*$ back into the square bias terms, and using \eqref{ry_dec} and \eqref{ry_dec2}, we obtain the upper bound for the rate of convergence of $\mathcal{R}_N(\widehat \lambda_j;h)$ and $\mathcal{R}_N(\widehat \psi_j;h)$.  
\end{proof}


\begin{proof}[Proof of Corollary \ref{corr_rates_est_cd}]
In the common design case, when $\mathfrak m^{2\underline H } / N \rightarrow \infty$, we have 
$(N\mathfrak m)^{2\underline H/(2\underline H+1)} /N \rightarrow \infty$. The risks of our estimators are then bounded by the risks on the infeasible, empirical estimators. When $\mathfrak m^{2\underline H } / N \rightarrow 0$, the squared bias term  remains larger than two others because the bandwidths cannot decrease faster than $m^{-1}$. In that case the risks' rate is bounded by 
$\{\log( \mathfrak m)/\mathfrak m\}^{2\underline H}$.
\end{proof}


\begin{proof}[Proof of Theorem \ref{thm:thm-1-bis}]
	Let us first notice that that the second part of Lemma A.\ref{lemma_Ih} 
 guarantees that the estimation of the function $H$ will not change the rates of convergence in probability.  
	Consider next the feasible versions of $\widehat \lambda_{j}$ and $\widehat \psi_{j}$ computed with all the unknown quantities replaced by estimates, as described in the Algorithm in Section \ref{sec:algorithm}. In particular, $\widehat \Gamma(s,t;h)$ is obtained with the diagonal correction	$\Hat{\Hat{d}}_N(s,t;h)$ in \eqref{diag_corr_eq_hat}.	
Theorem \ref{thm:thm-1-bis} is then a direct consequence of several facts. On the one hand, the uniform convergence of $\widehat \nu(t) $ and $\widehat c_2(s,t) $ from Step \ref{step3} of the Algorithm in Section \ref{sec:algorithm}. 
This uniform convergence follows from the uniform convergence of the presmoothing estimator $\widetilde X^{(i)}$. On the other hand, from the proof of Theorem \ref{thm:thm-1}, we have 
	$
	\widehat \lambda_j - \lambda_j = o_{\PP}(1)$ 
and
	$ \|\widehat \psi_j -  \psi_j \|_2 = o_{\PP}(1),
	$
	if $\widehat \lambda_j$ and $\widehat \psi_j$ are obtained from the Algorithm with  a constant function as  input proxy of $\psi_j$. 
\end{proof}

\vspace{-0.4cm} 

\section*{Acknowledgements}

The authors gratefully acknowledge funding from the French National Research Agency within the framework of the France 2030 programme for EUR DIGISPORT (ANR-18-EURE-0022) project.

\vspace{0.4cm} 
\noindent\textit{Conflict of interest}: None declared.

\vspace{-0.2cm} 

\section*{Data Availability}
\vspace{-0.1cm}

The computer code used in this article can be found at \url{https://github.com/sunnywang93/FDAdapt}. The real data analyzed  can be found at \url{https://archive.ics.uci.edu/dataset/235/individual+household+electric+power+consumption}. 

\vspace{-0.3cm} 

\section*{Supplementary Material}
In the Supplement, we provide detailed justification for the theoretical statements above. We also provide additional results from additional extensive simulation experiments with several different data generating processes. The content of the Supplement is the following. In Section \color{blue} I \color{black} we provide details on the properties of the class of multifractional Brownian motion introduced in Section \ref{sec:sim_design}. 
Section \color{blue} II \color{black}
is dedicated to the proofs of the properties of the local regularity and the proof of uniform convergence of our local regularity  estimators. Section \color{blue} III \color{black}
contain proofs of the technical results stated in the Appendix. In Section \color{blue} VI \color{black} 
we prove our first-order Taylor expansions for the eigen-elements, as previewed in Theorems A.\ref{th:theorem1_app}
and A.\ref{th:theorem2_app}
in the Appendix. Details for the proof of Theorem \ref{thm:thm-1} are given in Section \color{blue} V \color{black}. Sections \color{blue} VI \color{black} and \color{blue} VII \color{black}
contain additional materials on the case of differentiable sample paths, simulation  results and a real data analysis. 

\nocite{Belhakem2021}

\bibliographystyle{apalike}
\bibliography{references_KPW.bib}

\typeout{get arXiv to do 4 passes: Label(s) may have changed. Rerun}

\end{document}